\documentclass{article}
\usepackage{emulateapj}
\usepackage{apjfonts}

\begin{document}

\newcommand{\figtrka}{1}
\newcommand{\figimg}{2}
\newcommand{\figschem}{3}
\newcommand{\figspec}{4}
\newcommand{\figehb}{5}
\newcommand{\figtran}{6}
\newcommand{\figtrkb}{7}

\newcommand{\tabcat}{1}
\newcommand{\tabphot}{2}
\newcommand{\tablf}{3}
\newcommand{\tabcomp}{4}
\newcommand{\tabspur}{5}
\newcommand{\tabplum}{6}
\newcommand{\tabpagb}{7}
\newcommand{\tabmod}{8}

\submitted{To appear in The Astrophysical Journal}

\title{Detection and Photometry of Hot Horizontal Branch Stars in the
Core of M32$^{1}$}


\author{Thomas M. Brown$^2$, Charles W. Bowers, Randy A. Kimble,
Allen V. Sweigart}

\affil{Laboratory for Astronomy \& Solar Physics, Code 681, NASA/GSFC,
Greenbelt, MD 20771.   tbrown@pulsar.gsfc.nasa.gov,
bowers@band2.gsfc.nasa.gov, kimble@ccd.gsfc.nasa.gov, 
sweigart@bach.gsfc.nasa.gov.
}

\medskip

\author{Henry~C.~Ferguson}
\affil{Space Telescope Science Institute, 3700 San Martin Drive,
Baltimore, MD 21218.  ferguson@stsci.edu.}

\begin{abstract}

We present the deepest near-UV image of M32 to date, which for the
first time resolves hot horizontal branch (HB) stars in an elliptical
galaxy. Given the near-solar metallicity of M32, much larger than that
of globular clusters, the existence of an extended horizontal branch
is a striking example of the second parameter effect, and, most
importantly, provides direct evidence that hot HB stars and their
progeny are the major contributors to the UV upturn phenomenon
observed in elliptical galaxies. Our image, obtained with the Space
Telescope Imaging Spectrograph (STIS), detects approximately 8000
stars in a $25 \arcsec \times 25\arcsec$ field, centered $7.7\arcsec$
from the galaxy nucleus.  These stars span a range of 21--28 mag in the
STMAG system, and in the deepest parts of the image, our catalog is
reasonably complete ($>$~25\%) to a magnitude of 27.  The hot HB spans
a magnitude range of 25--27 mag at effective temperatures hotter than
8500 K.  We interpret this near-UV luminosity function with an
extensive set of HB and post-HB evolutionary tracks.

Although the UV-to-optical flux ratio in M32 is weak enough to be
explained solely by the presence of post-asymptotic giant branch
(post-AGB) stars, our image conclusively demonstrates that it arises
from a small fraction ($\lesssim$5\%) of the population passing
through the hot HB phase.  The production of these hot HB stars does
not appear to rely upon dynamical mechanisms -- mechanisms that may
play a role in the HB morphology of globular clusters.  The majority
of the population presumably evolves through the red HB and subsequent
post-AGB phases; however, we see far fewer UV-bright stars than
expected from the lifetimes of canonical hydrogen-burning low-mass
post-AGB tracks.  There are several possible explanations: (1) the
transition from AGB to T$_{\rm eff} > 60000$~K could be much more
rapid than previously thought; (2) the vast majority of the post-AGB
stars could be evolving along helium-burning tracks; (3) the post-AGB
stars could be surrounded by circumstellar dust during the transition
from the AGB to T$_{\rm eff} > 60000$~K.

\end{abstract}

\keywords{galaxies: evolution --- galaxies: abundances --- galaxies: stellar
content --- ultraviolet: galaxies --- ultraviolet: stars --- stars: evolution}

\section{INTRODUCTION} \label{secint}

The nearest elliptical galaxy available for study is M32, a companion
to M31.  Because M32 has a high surface brightness that is centrally 
concentrated, a near-solar metallicity, and a predominantly old
population, it is usually considered a ``compact elliptical'' galaxy,
as opposed to a ``dwarf elliptical'' or ``dwarf spheroidal'' (see Da
Costa 1997\markcite{D97}).  Due to its proximity (770 kpc),
space-based observations are able to resolve individual cool stars
near its center.  Hot stars can also be detected in the center itself,
because they are relatively rare, so crowding is not serious, and
because the cooler, dominant populations are suppressed in the
solar-blind UV bandpasses.  Because M32 can be studied through colors,
spectra, luminosity functions, and color-magnitude diagrams, it
represents a natural benchmark in our understanding of elliptical
galaxies.

Stellar population studies of M32 at optical and infrared wavelengths
have shown no evidence for an extended horizontal branch (HB)
(Grillmair et al.\ 1996\markcite{G96}), and population synthesis work
often assumes that the galaxy has a ``red clump'' HB\\ 

{\small $^1$Based on observations with the NASA/ESA Hubble Space
Telescope obtained at the Space Telescope Science Institute, which is
operated by the Association of Universities for Research in Astronomy,
Incorporated, under NASA contract NAS~5-26555.

$^2$ NOAO Research Associate.}\\
 
\noindent
morphology (e.g., Worthey 1994\markcite{W94}).  In the ultraviolet,
Brown et al.\ (1998\markcite{BFSD95}) found indirect evidence for an
extended HB in M32: the UV-bright post-HB stars apparently follow
evolutionary tracks originating from the hot end of the HB.  In
spectral synthesis studies, the assumption of a red clump HB
morphology contributes to the need for an intermediate-age component
to the stellar population, to provide the necessary ultraviolet flux
(see Grillmair et al.\ 1996\markcite{G96}
and references therein).  Direct proof of an extended HB would
therefore relax the requirement that M32 has a composite
population, with a dominant old component (of age 8 Gyr or more), and
a minority younger component (of age $\sim$5 Gyr).

Considered as an extreme case in the sequence of true elliptical
galaxies, M32 has the weakest UV-to-optical flux ratio measured to
date ($1550-V = 4.5$ mag; Burstein et al.\ 1988\markcite{B88}).  Known
as the UV upturn, the sharp rise in the spectra of elliptical
galaxies at wavelengths shorter than 2500~\AA\ was discovered with the
OAO-2 satellite (Code 1969\markcite{C69}).  Prior to this discovery,
researchers did not expect such a UV-bright component in supposedly
old, passively-evolving populations. Many candidates were suggested to
explain the upturn, including young massive stars, hot white dwarfs,
hot HB and post-HB stars, and non-thermal activity (for a complete
review, see Greggio \& Renzini 1990\markcite{GR90}; O'Connell
1999\markcite{O99}). As the measurements of local ellipticals were
expanded, IUE observations demonstrated a large variation in the
strength of the UV upturn from galaxy to galaxy, even though the
spectra of ellipticals appear very similar at longer wavelengths.
Characterized by the $1550-V$ color, the UV upturn becomes stronger
and bluer as the metallicity (optical Mg$_2$ index) of the galaxy
increases, while visible colors become redder (Burstein et al.\
1988\markcite{B88}).

Today, it is widely believed that HB stars and their progeny are
responsible for the far-UV emission in elliptical galaxies. There are
three classes of post-HB evolution, each evolving from a different
range of effective temperature on the zero-age HB (ZAHB; see Figure
\figtrka).  Following core helium exhaustion, stars on the red end of
the HB will evolve up the asymptotic giant branch (AGB), undergo
thermal pulses, evolve as bright post-AGB stars to hotter
temperatures, possibly form planetary nebulae, and eventually descend
the white dwarf (WD) cooling curve.  At hotter temperatures (and lower
envelope masses) on the HB, stars will follow post-early AGB
evolution: they evolve up the AGB, but leave the AGB before the
thermal pulsing phase, continue to high temperatures at high
luminosity, and descend the WD cooling curve.  The bluest HB stars,
with very little envelope mass, will follow AGB-Manqu$\acute{\rm e}$
evolution, evolving directly to hotter temperatures and brighter
luminosities without ever ascending the AGB, and finally descending
the WD cooling curve.  These three classes of post-HB behavior have
very different lifetimes, in the sense that the post-AGB stars are
bright in the UV for a brief period ($\sim 10^3 - 10^4$ yr), the
post-early AGB stars are UV-bright for a longer period ($\sim
10^4-10^5$ yr), and the AGB-Manqu$\acute{\rm e}$ stars are UV-bright
for very long periods ($\sim 10^6-10^7$ yr).  The HB phase itself
lasts $\sim 10^8$~yr, and so the presence of hot HB stars in a
population, combined with their long-lived UV-bright progeny, can
produce the strong UV upturn seen in the most massive, metal-rich
ellipticals (see Dorman, O'Connell, \& Rood 1995\markcite{DOR95} and
references therein).  In galaxies with a weak UV upturn, a significant
fraction of the far-UV flux can theoretically come from post-AGB stars
(see Brown et al.\ 1997\markcite{BFDD97}), and in the weakest UV
upturn galaxies -- such as M32 -- the spectra alone do not require the
presence of a hot HB.

The ZAHB is not only a sequence in effective temperature: it is also a
sequence in mass (see Dorman, Rood, \& O'Connell
1993\markcite{DRO93}).  For $\rm T_{eff} \gtrsim 6000$~K, a small
change in envelope mass ($\lesssim 0.1$~$M_\odot$) corresponds to a
large change in T$_{\rm eff}$ ($\Delta \rm T_{eff} \gtrsim 10,000$~K),
assuming solar abundances.  Because the main-sequence turnoff (MSTO)
mass decreases as age increases, the ZAHB will become bluer as a
population ages, assuming all other parameters (e.g., mass loss on the
red giant branch, metallicity, helium abundance, etc.)  remain fixed.
Note that this does not necessarily imply that age is the ``second
parameter'' of HB morphology (e.g., Fusi Pecci \& Bellazzini
1997\markcite{FB97}; VandenBerg 1999\markcite{V99}; Sweigart
1999\markcite{S99}).  The assumed first parameter of the HB morphology
debate is metallicity; the HB becomes bluer at lower metallicity,
assuming all other parameters (age, mass loss, etc.)  remain fixed.
This is due to two reasons.  First, the MSTO mass at a given age is
lower at lower metallicity, because a metal-poor star is more luminous
(and thus shorter-lived) than a metal-rich star of the same mass.
Second, as the metallicity decreases, the envelope opacity decreases,
leading to a higher effective temperature on the HB.  In short, HB
morphology tends to become bluer at lower metallicity and higher ages,
but other parameters may also play a role (rotation, He abundance,
deep mixing, etc.).  Although HB morphology in globular clusters (GCs)
tends to become redder at increasing metallicity, there are examples
of relatively metal-rich GCs with extended blue HBs (e.g., Rich et
al.\ 1997\markcite{R97}), leading to the ``second parameter'' debate.
Although elliptical galaxies should not be considered as overgrown
globular clusters, they are generally believed to be metal-rich, old
populations, and thus the presence of a hot HB in M32 can provide
further insight into the production of blue HBs.

M32 has been the subject of several UV imaging programs with the
Hubble Space Telescope (HST).  The earliest of these observations
(King et al.\ 1992\markcite{K92}; Bertola et al.\ 1995\markcite{B95})
were taken with the Faint Object Camera (FOC) before the correction of
the spherical aberration, and were subject to large photometric
uncertainties due to the calibration and aperture corrections.  After
the HST refurbishment with COSTAR, Brown et al.\
(1998\markcite{BFSD98}) also observed M32 with the FOC, and detected
the UV-bright post-HB phases of stars descended from the hot HB;
although the calibration was improved in these later observations, the
FOC was still subject to considerable uncertainties, such as an
unexplained format dependence to the photometric zero points, and red
leak.  Here, we describe the deep UV imaging of M32 performed with the
Space Telescope Imaging Spectrograph (STIS), a 2nd-generation HST
instrument with a vastly improved performance and calibration.  Our
image is sufficiently deep to reach the HB at temperatures hotter than
8500~K, and in fact does reveal the presence of these stars, providing
the first direct detection of hot HB stars in an elliptical galaxy.
We interpret our data with a new set of evolutionary tracks, and
compare our results to expectations from canonical tracks in the
literature.

\medskip
\parbox{3.0in}{\epsfxsize=3.5in \epsfbox{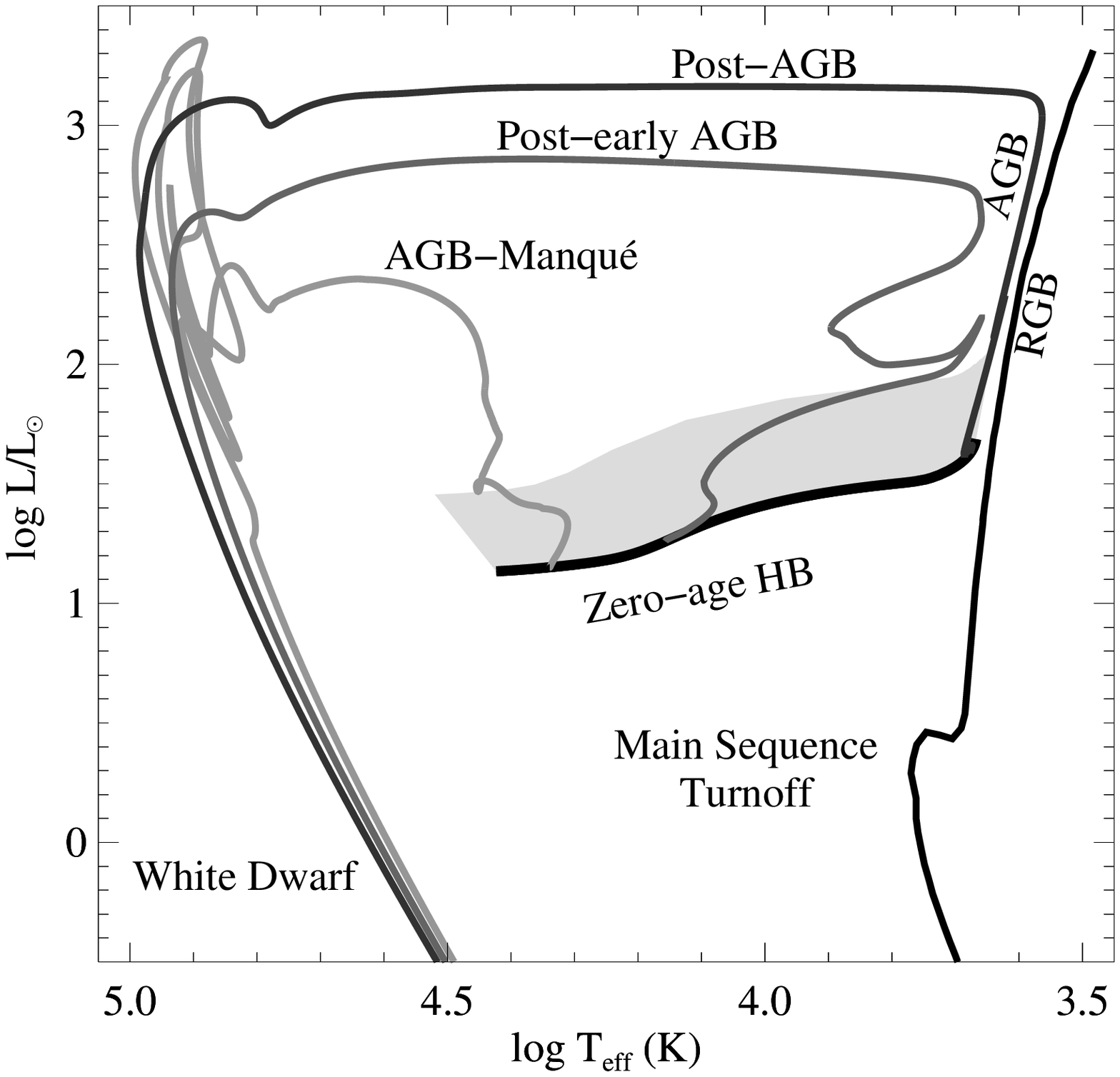}}
\centerline{\parbox{3.0in}{\small {\sc Fig.~\figtrka--}
The three classes of post-HB 
evolution arising from different ranges of effective
temperature and envelope mass on the zero-age HB.  The
tracks shown assume solar metallicity and helium abundance, and
are from our own calculations (see \S\ref{secmod}).  For populations
older than 1~Gyr, the HB phase (light grey)
is brighter than the main sequence turnoff, which
also contributes to the flux in the near-UV.
}}\vspace{0.2in}
\addtocounter{figure}{1}

\newpage

\hspace{0.75in}
\parbox{5.5in}{
{\sc Table \tabcat:} Near-UV Photometric Catalog (abridged)

\begin{tabular}{|r|r|r|r|r|r|r|}
\tableline
x & y & STMAG & error & region & RA\tablenotemark{a} & Dec\tablenotemark{a} \\
(pix) & (pix) & (mag) & (mag) & (\#) & (J2000) & (J2000) \\
\tableline
  38.6 &  112.9 & 25.64 &  0.11 & 1 &   0:42:40.4795  &  40:51:39.109\\
  39.0 &  578.3 & 26.41 &  0.15 & 1 &   0:42:41.3862  &  40:51:34.325\\
  39.2 &  199.4 & 27.33 &  0.40 & 1 &   0:42:40.6486  &  40:51:38.233\\
 ... & ... & ... & ... & ... & ... & ... \\
 354.9 &  821.3 & 22.39\tablenotemark{b} &  0.02 & 3 &   0:42:42.1502  &
  40:51:38.900\\
 355.8 &  620.1 & 26.44 &  0.24 & 3 &   0:42:41.7592  &  40:51:40.992\\
 ... & ... & ... & ... & ... & ... & ... \\
\tableline
\end{tabular}

$^a$Note that the relative astrometry is quite
accurate (tenths of a 0.025$\arcsec$ STIS pixel), but the absolute
astrometry is subject to a $\sim$1\arcsec --2$\arcsec$ uncertainty (associated
with the position of the guide stars).\\
$^b$Stars associated with planetary nebulae in our 
field.  Note that one of the four PNe in our field is at the edge of the
frame and is thus not included in the catalog.\\}

\section{OBSERVATIONS} \label{secobs}

STIS observed M32 on 19 October and 21 October 1998, with four
exposures on each day; the total exposure of the summed images is
22962 sec.  The combined image, shown with a logarithmic stretch to
enhance the faint stars, is shown in Figure \figimg.  The data were
taken with the near-UV multi-anode microchannel array (NUVMAMA), using
the crystal quartz
filter (F25QTZ); the filter blocks light shortward
of 1450~\AA\ and thus reduces the sky background from terrestrial
airglow lines of \ion{O}{1} and Lyman $\alpha$.  The long wavelength
cutoff of the bandpass is 3500~\AA, due to the sensitivity of the
detector, and red leak at longer wavelengths is minimal (Baum et al.\
1998\markcite{B98}).  The STIS MAMAs are photon counters that register
less than one count per incident cosmic ray; thus, cosmic ray 
rejection is not required, as it is with CCD imaging 
(where an incident cosmic ray causes a massive many-count ``hit'').
A full description of the instrument and
its capabilities can be found in Woodgate et al.\ (1998\markcite{W98})
and Kimble et al.\ (1998\markcite{K98}).

\begin{figure*}
\parbox{7.0in}{\epsfxsize=7.0in \epsfbox{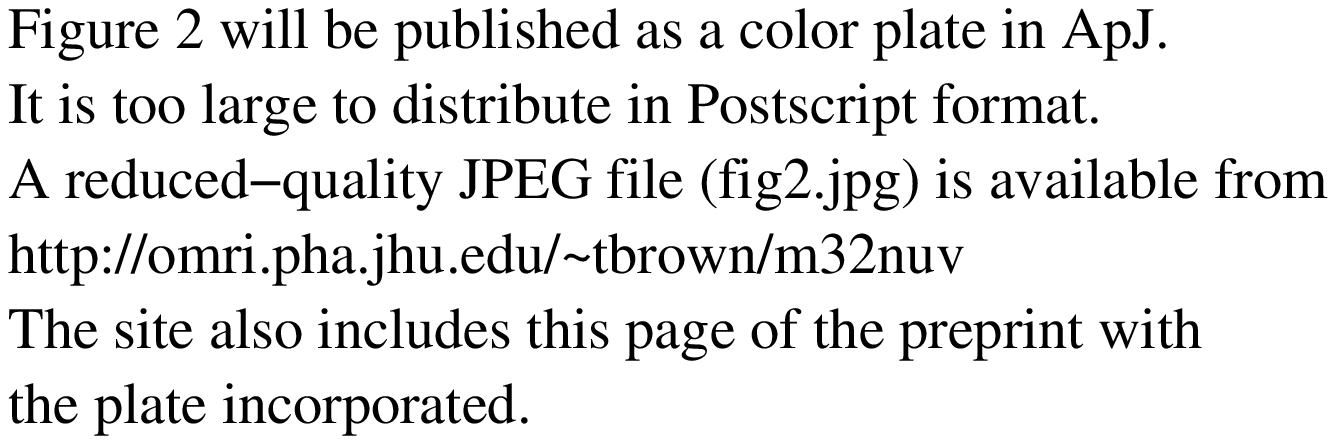}}
\caption{ This false-color image of the M32 center uses a logarithmic
scaling to enhance the faintest stars on the HB.  The STIS field is
$25 \arcsec \times 25 \arcsec$, and the combination of the dithered
positions have been summed onto a $1060\times 1060$ pixel image (cropped
to $1024 \times 1024$ pixels here).
The occulted portions of the image are due to a scattered light mask, 
and vignetting is negligible.
}
\end{figure*}

The $25\arcsec \times 25\arcsec$ ($1024 \times 1024$ pixel) field was
centered 7.7$\arcsec$ 
south of the M32 center, in order to overlap with earlier FOC UV imaging
of M32 (Brown et al.\ 1998\markcite{BFSD98}), to allow imaging in
regions of lower diffuse galaxy background while still including the
center of the galaxy, and to place the field farther from M31.  The
images were dithered between three positions along the X/Y diagonal of
the detector: $+0.23\arcsec/+0.23\arcsec$, $0\arcsec/0\arcsec$,
and $-0.23\arcsec/-0.23\arcsec$.  Given the plate scale of 24.465
mas pix$^{-1}$, these offsets were equivalent to shifts of $\approx$9
pixels along each axis of the detector, which is useful for smoothing
out small-scale variations in detector sensitivity.  A small fraction
of the field near the detector edge is occulted by a scattered-light
mask very close to the focal plane: the lower right corner
($\approx$9000 pixels), the upper right corner ($\approx$2700 pixels),
the bottom 20 rows, the first 8 columns, the top row, and last 4
columns.  Although the mask slightly decreases the field of view, it
does not vignette stars in the remainder of the field, and it does
allow an accurate characterization of the dark counts in the data.

The STIS focal plane is tilted with respect to the detector when in
imaging mode (STIS is optimized for spectroscopy).  Thus, the PSF
changes in shape and width as one moves from left to right in the
field, and it also changes in time due to telescope breathing.  Our
final summed image averages over time to produce a PSF that is very
nearly circular in the left half of the image, and noticeably
elliptical in the right half of the image.  Our detection limits are
driven by the combination of both focus and galaxy background.  In
this image, the focus is best where the diffuse background from the
core of M32 is faintest.  This combination makes the limiting
sensitivity significantly
\\ \vskip 2.15in \noindent 
non-uniform over the summed image, but
provides the deepest possible sensitivity in the well-focussed
portions of the image.  An elliptical Gaussian fit to isolated bright
stars in the field has a FWHM range of 2.42--3.96 pix
(0.06--0.1$\arcsec$), and an axial ratio range of 0.64--0.95.

We detect $\sim$8000 stars in the near-UV.  Because the image contains
$\approx 2.4\times10^5$ resolution elements (in the regions where we
are actually cataloging stars), this is obviously a crowded field,
with an average of one star per 30 resolution elements; near the M32
center, the crowding is severe, with one star per 12 resolution
elements, but farther from the center it drops to one star per 63
resolution elements.  The combination of this crowding with the
variable PSF and the variable galaxy background makes photometry
difficult, but not impossible, as we will discuss below.

Although the internal interstellar extinction within M32 is unknown,
it is presumably small in comparison to the Galactic foreground
extinction, thought to be anywhere in the range $0.035 \le E(B-V) \le
0.11$ mag (Tully 1988\markcite{Tu88}; McClure \& Racine
1969\markcite{MR69}; Burstein et al.\ 1988\markcite{B88}; Ferguson \&
Davidsen 1993\markcite{FD93}; Burstein \& Heiles 1984\markcite{BH84}).
For the purposes of this paper, we assume $E(B-V) = 0.11$~mag when
comparing the data to the predictions of stellar evolutionary tracks,
corresponding to 0.8 mag of extinction in the STIS bandpass.

Of the $\sim 8000$ stars resolved in the STIS image, four of the
brightest roughly coincide with the positions of the only known
planetary nebulae (Ciardullo et al.\ 1989\protect\markcite{C89}) in
this field.  Specifically, if one takes their positions in the SIMBAD
database, there is a very bright star in the STIS image approximately
1$\arcsec$ west of each position, implying that there is a global
offset between the astrometry of our image and that of the planetary
nebulae.  These stars are flagged in our catalog (Table \tabcat,
available from the CDS), and their positions are marked in
Fig~\figschem.

\section{DATA REDUCTION} \label{secred}

We reduced the raw images via the CALSTIS package in IRAF, including
an updated pixel-to-pixel flat field file.  Two steps in the reduction
were done outside of CALSTIS: geometric correction was applied during
the final combination via the IRAF DRIZZLE package, and dark
subtraction was performed by subtracting a scaled and flat-fielded
dark image from each data frame, with the scaling determined from the
occulted right-hand corners of the detector.  All frames were
cross-correlated to provide accurate relative shifts.  The calibrated
frames were then drizzled to a $1060 \times 1060$ pixel image that
included the data from all dither positions.
The resulting image has a narrow strip of underexposed pixels along
the edges, due to the dither

\hspace{0.5in}
\parbox{6.25in}{
{\sc Table \tabphot:} Photometry

\begin{tabular}{|c|r|r|r|r|r|r|}
\tableline
       & 
x-position$^a$ & 
FWHM$^b$ & 
resolution & 
detected & 
fraction of near-UV & 
resolved fraction   \\
region & 
(pix)  & 
(pix)  & 
elements & 
stars  & 
flux resolved$^c$ & 
$\ge 25.5$~mag$^d$ \\

\tableline
1      & 37--141  & 2.68 & 30806 &  556 & 0.33 & 0.17 \\
2      & 142--341 & 2.52 & 63677 & 1008 & 0.19 & 0.11 \\
3      & 342--541 & 2.62 & 59818 & 1316 & 0.13 & 0.09 \\
4      & 542--741 & 3.00 & 44864 & 2150 & 0.11 & 0.10 \\
5      & 742--941 & 3.43 & 27366 & 2226 & 0.10 & 0.11 \\
6      & 942--1037& 3.62 & 13035 &  765 & 0.09 & 0.10 \\
\tableline
\end{tabular}

$^a$Each region extends the full height of the image, excluding
occulted detector areas and a $4 \arcsec \times 2.7 \arcsec$ ellipse centered 
on M32.  The galaxy center is at $(x,y) = (842.7,392.7)$.\\
$^b$This is the average FWHM of a Gaussian fit to the stars 
in each region, and is used for object detection.\\
$^c$From all detected stars in the uncorrected catalog, and assuming a sky 
background of $9 \times 10^{-4}$ cts sec$^{-1}$ pix$^{-1}$.\\
$^d$From the corrected luminosity function, assuming a sky 
background of $9 \times 10^{-4}$ cts sec$^{-1}$ pix$^{-1}$.\\
}\\ \\ \\

\noindent
pattern.  This strip is scaled correctly
to account for the smaller exposure time, but thus has higher noise
than the fully exposed area of the image.  Stars in the underexposed
strip are not included in our catalog, and this strip was not included
in the Monte Carlo simulations discussed in \S\ref{seccom} and
\S\ref{secspur}.

The drizzle ``dropsize'' (also known as pixfrac) was 0.1, thus
improving the resolution over a dropsize of 1.0 (which would be
equivalent to simple shift-and-add).  This small dropsize does not
cause problematic ``holes'' in the final image, because the pixel
scale was unchanged; it is thus equivalent to interlacing the
individual frames. In the final image, pixels outside of the dither
pattern or occulted are set to a count rate of zero.  The pixel mask
used in the drizzle for each input frame included the occulted regions
of the detector, a small number of hot pixels, and pixels with
relatively low response (those with values $\le 0.75$ in the 
pixel-to-pixel flat field).

\section{PHOTOMETRY} \label{secphot}

\subsection{PSF Fitting} \label{secpsf}

Because the M32 image is fairly crowded, there are two obvious options
for performing photometry: PSF fitting or small-aperture photometry.
However, the PSF also varies strongly with position in the image, and
so the aperture correction for small-aperture photometry would be a
strong function of position, possibly introducing systematic errors in
our catalog.  We felt that PSF fitting offered the most accuracy for
these data. As an initial estimate of the PSF variation, we used the
IRAF routine FITPSF to fit an elliptical Gaussian to 180 of the
brightest isolated stars in the image.  The result defined a map of
the Gaussian semi-major axis width as a function of position in the
image.  The map confirmed that the PSF varies as a function of
horizontal position.

For our PSF fitting, we used the January 1998 version of the
DAOPHOT-II package (Stetson 1987\markcite{S87}).  The software allows
calculation of relative photometry with a variable PSF defined by the
user.  Our iterative procedure was similar to that described in the
User's Manual, to which we refer interested readers for details.
However, because the PSF and the diffuse background from unresolved
stars both vary with position, we chose to perform our photometry on
six image regions (defined in Table \tabphot).  Object detection and
PSF fitting for each region were done with the region boundaries
extended 50 pixels into the neighboring regions, to avoid edge effects
with stars near the region boundaries.  Thus, at each region boundary,
there is a 100-pixel wide strip where stars are fit twice, and we use
this \\ \vskip 2.4in 

\hskip -0.2in
\parbox{3.0in}{\epsfxsize=3.5in \epsfbox{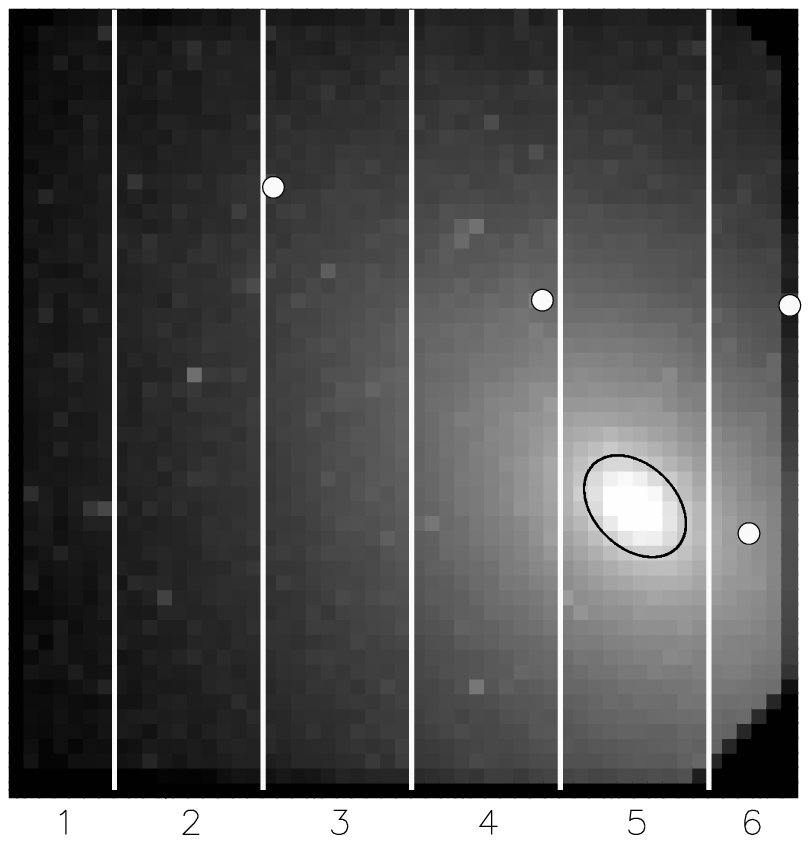}} \vskip 0.1in
\centerline{\parbox{3.0in}{\small {\sc Fig.~\figschem--}
This schematic of the STIS image shows some of the details of the
reduction discussed in the text.  The six labeled vertical strips are
the regions used for PSF fitting and photometry.  Stars were rejected
from the catalog within the black ellipse in the center of the galaxy.
White circles show the positions of the bright stars
that correspond to the positions of the four known planetary nebulae in
our field; from left to right, they are: Ford-NGC221-21, Ford-NGC221-23, 
Ford-NGC221-27, and Ford-NGC221-24.
}}
\addtocounter{figure}{1}
\smallskip

\noindent
overlap to put the stars in each region on the same magnitude
system (see below).  Each region spanned the entire height of the
image, with region 5 encompassing the center of the galaxy.  From our
initial map of elliptical Gaussian fits, we determined the ``average''
width of a star in each region (Table \tabphot), and used this width
for object detection.  DAOPHOT convolves an image with a Gaussian as
part of its object detection algorithm, but the user must specify a
constant width for each image under consideration.  Thus, our use of a
different width for each region allows optimization of the object
detection.

To determine a variable PSF for each region, we first subtracted a
median-filtered image from the data to remove the varying galaxy
background.  We found that this subtraction prevented a spurious
biasing of the PSF wings; fitting the PSF with the galaxy background
intact incorrectly tilts the PSF wings along the slope of the galaxy
background.  Our PSF in each region is a Moffat function
with a lookup table of empirical corrections.  The PSF in each region
was allowed to vary linearly with position.  Although DAOPHOT allows
quadratic variations in the PSF, our PSF subtraction was worsened when
we attempted to allow such higher order variation.  The PSF
determination was done though an interactive selection of the
brightest isolated stars, with iterations to subtract the PSFs of the
neighboring stars, as described in the manual for the software.  The
PSF was defined with a radius of 10 pixels.

After determining a variable PSF for each image region, we performed
object detection and PSF fitting on the data, again using the regions
defined in Table \tabphot\ (but without the galaxy background
subtraction).  To maintain positional consistency, we did not actually
extract the regions from the image; we simply set those pixels outside
of a given region to a ``bad'' pixel value, thus producing a mask for
each region.  In region 5, which contains the M32 center, we also masked
off a $0.75 \arcsec \times 0.5 \arcsec$ ellipse aligned to the axis of
M32 (45$^{\rm o}$ counterclockwise from the +y axis of the image), in
order to prevent an unnecessary biasing of the ``sky'' level, and
because the very center appeared hopelessly crowded.  In the final
catalog, we actually rejected stars in a larger $4 \arcsec \times 2.7
\arcsec$ ellipse, after reevaluating the severity of the crowding.  We
set the detection threshold to 5~$\sigma$; setting a lower threshold
produces many false detections, because the diffuse background varies
with position, even within the subsections we defined.  The detection
algorithm convolves the image with a symmetrical Gaussian with a width
equal to the FWHM of the PSF, and then looks for deviations above the
local sky background (see Stetson 1987\markcite{S87} for a complete
description).  We set the ``roundness'' and ``sharpness'' criteria to
twice their default values, because the STIS PSF is not symmetrical
and there is no need to reject cosmic rays in STIS MAMA images.  After
a first pass of object detection and PSF fitting, the star-subtracted
image is used for a second pass at object detection, and the final
combined catalog is used for a second pass of PSF fitting.  The PSFs
were fit to the stars using a fitting radius of 3 pixels and a sky
annulus spanning 8--15 pixels in radius.  The DAOPHOT package is
capable of iteratively solving for the sky under the star itself
(hence the inner radius of the sky annulus falls within the PSF
radius).

DAOPHOT calculates the {\it relative} magnitudes of stars in each
region.  Putting all of the stars in each region on the same system
required several additional steps.  The PSF in region 2 is the most
circular and has the smallest FWHM.  We performed aperture photometry
on the 10 brightest stars in region 2, with the aperture varying from
3--10 pixels in radius.  Next, we performed the same aperture
photometry on the quasar in the Hubble Deep Field South (HDF-S)
near-UV image (Gardner et al.\ 1999\markcite{G99}).  The
encircled-energy variation with aperture size in our M32 image was
nearly identical to that of the quasar.  We then used the quasar (with
object masking for the other galaxies in the HDF-S) to determine an
aperture correction of 0.524 mag for a 3-pixel radius aperture, and
used this aperture correction to calculate the corrected magnitudes of
the 10 bright stars in M32.  This then determined the zero-point for
the DAOPHOT magnitudes in region 2.  Next, we used the 100-pixel-wide
overlaps at region boundaries (see above) to place all of the stars on
the same magnitude system as region 2; the photometry in each region
was shifted by less than 0.1 mag as part of this correction.  Our
magnitudes are specified in the STMAG system:\\
\[m = {\rm -2.5 \times log_{10}} f_{\lambda} -21.10\]
\[f_{\lambda} ={\rm counts \times PHOTFLAM / EXPTIME}\]

\noindent
where EXPTIME is the exposure time, and PHOTFLAM is 
$5.588 \times 10^{-18}$ erg s$^{-1}$ cm$^2$ \AA$^{-1}$ / (cts s$^{-1}$).

The entire photometric catalog (Table \tabcat) is available from the
CDS. Table \tablf\ shows the raw (uncorrected) luminosity functions in
each region.  \\

\begin{center}
{\sc Table \tablf:} Uncorrected Luminosity Functions

\begin{tabular}{|r|r|r|r|r|r|r|}
\tableline
 & \multicolumn{6}{|c|}{No. of stars in region}\\
\tableline
mag         & 1    & 2    & 3    & 4    & 5    & 6   \\
\tableline
21.0        & 0    & 0    & 0    & 0    & 1    & 0   \\
21.5        & 0    & 1    & 0    & 2    & 2    & 0   \\
22.0        & 0    & 0    & 1    & 4    & 5    & 2   \\
22.5        & 1    & 2    & 2    & 4    & 1    & 0   \\
23.0        & 2    & 1    & 2    & 7    & 6    & 1   \\
23.5        & 1    & 3    & 4    &10    &16    & 5   \\
24.0        & 6    & 9    & 12   &29    &69    & 16  \\
24.5        & 7    & 7    & 31   &58    &197   & 31  \\
25.0        &14    &43    & 95   &248   &403   & 146 \\
25.5        &66    & 196  & 328  &588   &649   & 271 \\
26.0        &110   & 306  & 416  &706   &555   & 214 \\
26.5        &154   & 255  & 323  &388   &254   & 69  \\
27.0        &139   & 135  &  92  &102   & 67   & 10  \\
$\ge$ 27.5  &56    &50    &  10  &  4   &  1   &  0  \\
\tableline
\end{tabular}
\end{center} 

\medskip

\subsection{Completeness} \label{seccom}

In the deepest regions of the image, we are detecting stars down to 28
mag.  However, the catalog becomes seriously incomplete well before
this point is reached, and the completeness varies from region to
region.  To determine the completeness as a function of magnitude in
each region, we ran thousands of Monte Carlo simulations, using the
DAOPHOT package and our own IDL-based programs.  For each region, 10
stars at a given magnitude were placed within the region at random
positions, using the variable PSF for that region, and including
Poisson noise in the artificial stars.  The entire object detection
and PSF fitting process was then performed, and the catalog was
checked to see if the stars were recovered.  This process was then
repeated until the calculated completeness was deemed well-determined,
with the criteria that the calculated completeness not change
significantly (by $>$1\%) over many runs.  Because we only add 10
stars per simulation, we are not significantly affecting the crowding
in the data (which contains hundreds of stars in each region).
Table \tabcomp\ lists the completeness versus magnitude for each
region.  Note that the object detection is done with a Gaussian
convolution of constant width (specified in Table \tabphot),
while the artificial stars are created using the empirically-corrected
Moffat function that was determined from the STIS data (and
significantly different from a Gaussian); thus, the completeness
determination is not a circular process, and it does accurately model
the true detection process.  Because the strength of the galaxy background
has a strong influence on the completeness, the completeness is best in
region 1, even though the PSF is sharpest in region 2.

To ensure that inaccuracy in our variable PSF template does not skew
the determination of the completeness, we recalculated a set of
completeness simulations using smoothed PSFs for the artificial stars.
Because these test PSFs were not as strongly peaked, they could
possibly reduce the detectability.  We first smoothed the true PSF
with a kernel that redistributed 10\% of the flux in each pixel to its
8 neighbors, and then with a kernel that redistributed 20\% of the
flux in each pixel to its 8 neighbors.  The artificial stars were then
added to the data with Poisson noise, as before.  Although we used
these {\it altered} PSFs to generate the artificial stars in the Monte
Carlo simulations, we used the {\it true} PSF for the PSF fitting,
thus simulating a mismatch between the object PSF and the fitted PSF.
Fortunately, the recalculated completeness versus magnitude remained
virtually unchanged when these altered PSFs were used, thus
demonstrating that small errors in the assumed PSF do not
significantly affect the completeness determination.

\medskip

\begin{center}
{\sc Table \tabcomp:} Completeness in Near-UV Number Counts

\begin{tabular}{|c|c|c|c|c|c|c|}
\tableline
 & \multicolumn{6}{|c|}{region}\\
\tableline
mag  & 1    & 2    & 3    & 4    & 5    & 6   \\
\tableline
21.5 & 1.00 & 1.00 & 1.00 & 1.00 & 0.91 & 0.98\\
22.0 & 1.00 & 1.00 & 1.00 & 1.00 & 0.90 & 0.97\\
22.5 & 0.99 & 0.99 & 1.00 & 0.98 & 0.88 & 0.96\\
23.0 & 1.00 & 0.99 & 0.99 & 0.98 & 0.89 & 0.96\\
23.5 & 0.99 & 0.99 & 0.98 & 0.96 & 0.85 & 0.94\\
24.0 & 0.99 & 0.98 & 0.98 & 0.95 & 0.83 & 0.92\\
24.5 & 0.97 & 0.97 & 0.96 & 0.93 & 0.81 & 0.89\\
25.0 & 0.97 & 0.96 & 0.95 & 0.87 & 0.73 & 0.77\\
25.5 & 0.96 & 0.96 & 0.92 & 0.79 & 0.55 & 0.60\\
26.0 & 0.92 & 0.87 & 0.71 & 0.48 & 0.29 & 0.22\\
26.5 & 0.68 & 0.46 & 0.25 & 0.14 & 0.08 & 0.04\\
27.0 & 0.24 & 0.11 & 0.05 & 0.04 & 0.02 & 0.01\\
27.5 & 0.06 & 0.02 & 0.01 & 0.00 & 0.01 & 0.00\\
\tableline
\end{tabular}
\end{center}

\medskip

\subsection{Spurious Sources} \label{secspur}

Because we set the DAOPHOT detection threshold a bit higher than the
nominal value of 4~$\sigma$, we do not expect many spurious sources to
be detected in the data, except perhaps in region 5, where the galaxy
background varies strongly with position.  However, to quantify
exactly how many spurious detections we have in our catalog, we ran
another series of Monte Carlo simulations.

Each simulation starts with a model of the diffuse light, derived by
taking the median of the true image, and assuming elliptical isophotes
to handle edge effects and occulted regions.  This galaxy model was
then used to create a starless simulation of each of the eight STIS
exposures, by dithering the model, scaling by the exposure time,
applying detector occultation in the corners and along the edges,
adding dark counts, and adding Poisson noise.  The eight simulated
exposures were then drizzled into one image, in the same manner as the
true data.  This image matched the noise characteristics of the actual
data in the areas free of point sources.  Finally, two passes of
object detection and PSF fitting were then applied, again following
the actual data reduction described above.  The process was repeated
100 times for each region, to determine the average number of spurious
sources one would expect as a function of magnitude in each region.
The results are tabulated in Table \tabspur.

As expected, spurious sources do not significantly contaminate our
catalogs where the completeness is larger than 10\%.  The number of
spurious sources also begins to drop at the faintest magnitudes,
because such stars fluctuate below the 5~$\sigma$ detection
limits.

\begin{center}
{\sc Table \tabspur:} Spurious Source Contamination

\begin{tabular}{|c|r|r|r|r|r|r|}
\tableline
 & \multicolumn{6}{|c|}{region}\\
\tableline
mag  & 1    & 2    & 3    & 4    & 5    & 6   \\
\tableline
23.5 & 0    & 0    & 0    & 0    & 0    & 0   \\
24.0 & 0    & 0    & 0    & 0    & 1    & 0   \\
24.5 & 0    & 0    & 0    & 0    & 3    & 0   \\
25.0 & 0    & 0    & 0    & 0    &14    & 0   \\
25.5 & 0    & 0    & 0    & 1    &42    & 1   \\
26.0 & 0    & 0    & 0    & 9    &88    &15   \\
26.5 & 0    & 0    & 1    &62    &120   &41   \\
27.0 & 3    & 6    &20    &55    &36    & 9   \\
27.5 &17    &21    &12    & 1    & 0    & 0   \\
\tableline
\end{tabular}
\end{center}

\subsection{Systematic Errors} \label{secsys}

\subsubsection{Comparison with FOC} \label{secfoc}

UV imaging of the evolved stellar populations in M31 and M32 has been
subject to considerable calibration problems, and the story of these
problems underscores the need for faint UV standards appropriate for
direct imaging with today's sensitive instruments.  King et al.\
(1992\markcite{K92}) imaged M31 and M32 with the pre-COSTAR FOC and
assumed that the FOC sensitivity was degraded to 80\% of nominal.
Bertola et al.\ (1995\markcite{B95}) imaged the same galaxies, and
their adoption of a degraded FOC sensitivity (at 30--40\% of nominal in
the UV) implied even brighter UV luminosities for these stars.  Both
sets of pre-COSTAR data required large aperture corrections ($>$ 2
mag) because of the spherical aberration present in the
first-generation instruments.  Subsequent recalibration of the FOC
showed that the UV sensitivity of the King et al.\
(1992\markcite{K92}) data was supposedly at 144\% of nominal, due to
the format dependence of the FOC zero-points (Greenfield
1994\markcite{G94}).  After refurbishment of HST, Brown et al.\
(1998\markcite{BFSD98}) used the FOC to image M31 and M32, and found
that its nominal calibration implied that the stars common to both the
Brown et al.\ (1998\markcite{BFSD98}) data and the King et al.\
(1992\markcite{K92}) data were 1.9 mag fainter than previously thought
by King et al.\ (1992\markcite{K92}); this discrepancy dropped to 1.2
mag once the format dependence of the FOC zero-points was included in
the King et al.\ (1992\markcite{K92}) magnitudes.  Brown et al.\
(1998\markcite{BFSD98}) attempted several cross-checks of the
refurbished FOC calibration, and found inconclusive evidence for 0.25
mag systematic shifts to the zero points, in the sense that these
UV-bright stars could be 0.25 mag brighter than implied by the nominal
calibration of the refurbished FOC.  However, these checks were
done via comparison to spectra and galaxy photometry -- no images
of globular clusters or isolated stars were available in the appropriate
imaging modes.

Our new STIS observations of M32 include the entire field observed by
Brown et al.\ (1998\markcite{BFSD98}) with the FOC.  Given the STIS
magnitudes ($m_{NUV}$) and the FOC magnitudes ($m_{F175W}$ and
$m_{F275W}$) for these stars, color-color diagrams of $m_{NUV} -
m_{F175W}$ versus $m_{F175W} - m_{F275W}$ and $m_{NUV} - m_{F275W}$
versus $m_{F175W} - m_{F275W}$ demonstrate that the FOC magnitudes in
Brown et al.\ (1998\markcite{BFSD98}) should be revised 0.5 mag
brighter than listed in the Brown et al.\ (1998\markcite{BFSD98})
catalog, assuming that the observed stars span a range of effective
temperature from 8000--30000~K.  This revision would further reduce
the discrepancy between the Brown et al.\ (1998\markcite{BFSD98}) data
and the King et al.\ data (1992\markcite{K92}) to 0.7 mag, and so this
remaining discrepancy might be due to the large aperture corrections
in the pre-COSTAR data.  However, we note that this means the
UV-bright stars in the King et al.\ (1992\markcite{K92}) data are 1.4
mag fainter than originally thought by King et al.\
(1992\markcite{K92}); thus, these stars are not bright enough to be
post-AGB stars (the interpretation of King et al.\ 1992\markcite{K92}
and Bertola et al.\ 1995\markcite{B95}), and are consistent with
post-HB evolution from the hot end of the HB, as described in Brown et
al.\ (1998\markcite{BFSD98}).  Shifting the stars in the Brown et al.\
(1998\markcite{BFSD98}) color-magnitude diagrams by 0.5~mag in both
filters does not move the bulk of the stars onto post-AGB tracks.

\subsubsection{Comparison with WFPC2} \label{secwfpc}

The STIS photometric calibration is reliable at the 0.15 mag level,
according to the STScI documentation (Baum et al.\
1998\markcite{B98}); this is considerably more secure than that of
the earlier generation instruments on the HST.  At the time of this
writing, new calibration efforts at STScI will revise the photometric
zero-points slightly, and adaptation of these new zero points would
make our stars up to 0.07~mag brighter.  Our analysis is not
sensitive to such a small shift in sensitivity, and so we do not adopt
the revision. 

Our own consistency check, using STIS \& WFPC2 UV images of the
globular cluster NGC6681, confirms the photometric accuracy of STIS. In the
UV, this GC is not crowded and does not have an underlying galactic
background; it gives a much better sensitivity check than the data
available for checks of the FOC calibration.  STIS imaged NGC6681 with
the near-UV crystal quartz filter and the far-UV SrF$_2$ filter
($m_{FUV}$); 22 bright isolated stars are available in the field for
accurate photometry.  The same field was also observed by WFPC2 using
the F160BW filter ($m_{F160BW}$).  The F160BW filter has very low
throughput, but very little red leak.  These bright stars have
$\lesssim 0.05$ mag statistical errors in the WFPC2 frame, and
$\lesssim 0.01$ mag statistical errors in the STIS frames.
Color-color diagrams of $m_{FUV} - m_{F160BW} $ versus $m_{FUV} -
m_{NUV}$ and $m_{NUV} - m_{F160BW} $ versus $m_{FUV} - m_{NUV}$ are
consistent at the 0.1 mag level when compared to expectations for
stars spanning a range of effective temperature 8000--30000~K.  The
expected colors were calculated by folding the synthetic spectra of
Kurucz (1993) through the bandpasses of the IRAF SYNPHOT package.

\subsubsection{Comparison with IUE} \label{seciue}

The center of M32 was observed by IUE at low signal-to-noise (S/N) in
the wavelength range 1150--3200~\AA, thus covering most of the STIS
bandpass.  We compare the STIS image to a composite UV+optical
aperture-matched spectrum (Calzetti, private communication); because
the red leak in STIS is so low, the optical portion of the spectrum
has little affect on the analysis, but is included for completeness.
We show our bandpass and this spectrum in Figure \figspec.

The IUE aperture was a $10 \arcsec \times 20 \arcsec$ oval.  Because
M32 is not centered in the STIS image, we must define two regions in
the STIS field that can be used for comparison to IUE: a $10 \arcsec
\times 10 \arcsec$ square centered on the M32 nucleus, and an adjacent
semicircle with a radius of $5 \arcsec$.  The addition of the flux in
the square with twice that of the semicircle is equivalent to the IUE
aperture.

Within this artificial aperture, STIS measures a dark-subtracted count
rate of 3228 cts sec$^{-1}$.  A template of the ``average'' sky
background, available from the STScI web page, shows that $\sim$275
cts sec$^{-1}$ of this could be from the sky background, and so the
sky-subtracted count rate measured by STIS in an IUE-equivalent
aperture is $\sim$2950 cts sec$^{-1}$.  Folding the composite
IUE+optical spectrum of IUE through the STIS bandpass (via the IRAF
package SYNPHOT) predicts 2738 cts sec$^{-1}$, within 10\% of the
actual value measured by STIS.  Given the uncertainty in the true sky
level, the IUE and STIS observations are in acceptable agreement.

\medskip
\hskip -0.15in
\parbox{3.5in}{\epsfxsize=3.5in \epsfbox{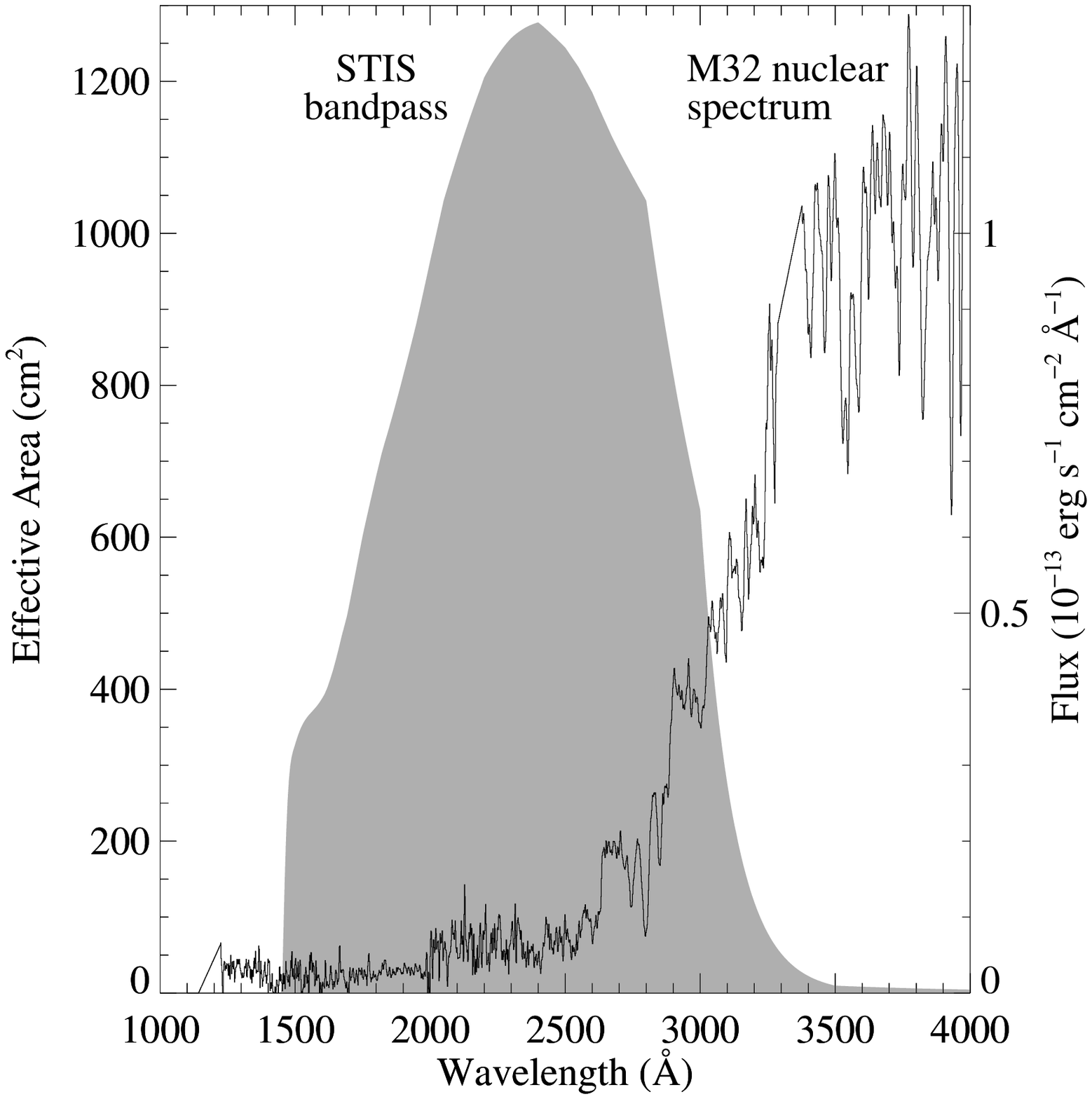}}
\centerline{\parbox{3.0in}{\small {\sc Fig.~\figspec--}
The nuclear ($10\arcsec \times 20\arcsec$) spectrum (solid; Calzetti,
private communication), an aperture-matched splice of IUE
and ground-based data, demonstrates that the STIS
bandpass (grey shaded) incorporates flux from both the short-wavelength
UV upturn population and the cooler stars in earlier evolutionary phases.
Given that these cooler populations are several magnitudes fainter than
our detection limit in this bandpass, they contribute to the diffuse
background in our data.}}
\addtocounter{figure}{1}

\section{RESOLVED FRACTION OF NEAR-UV LIGHT} \label{secres}

Because the STIS image is the deepest near-UV image of M32 to date, we
compare the resolved flux with the total flux in the STIS bandpass.
The fraction of resolved flux varies with position in the image: a
much larger fraction of the near-UV light is resolved into stars as
one moves away from the M32 center.  In Table \tabphot, we show the fraction 
of M32 flux within the STIS bandpass resolved into stars, for each of the
6 regions defined in the table.  In each region, we subtract a sky
background of $9 \times 10^{-4}$ cts sec$^{-1}$ pix$^{-1}$, which
comes from the ``average'' sky template available on the STScI web
page, folded through the SYNPHOT package bandpass.  We then calculated
the resolved fraction via two different methods.  In the first method,
we simply took the catalog of stars for that region, summed the flux
from those stars, and divided by the total flux in that region.  This
did not assume any completeness correction, nor did it reach a uniform
depth as a function of position in the image. In the second method, we
summed the flux in the corrected luminosity function for those
magnitude bins $\ge$ 25.5 mag (i.e., better than 50\% completeness
everywhere), and then divided by the total flux in that region.  Thus,
this second method reaches a uniform limiting magnitude across the
image.  We resolve $\sim$10\% of the near-UV flux in most of the
image, and about a third of the near-UV flux in the deepest section.
We do not expect to resolve all of the flux, because a considerable
fraction of the light in our bandpass comes from the cooler
populations that are below our detection limits in the near-UV (see
Figure \figspec).  It would take {\it much} deeper imaging in this
bandpass, at much higher resolution, to resolve significantly more
flux, because the main sequence turnoff is several magnitudes below
the HB (see Figure 1).

\section{COMPARISON WITH EVOLUTIONARY TRACKS} \label{secevo}

\subsection{The Stellar Evolutionary Flux} \label{secsef}

Before comparing our luminosity functions to the expectations from
stellar evolutionary tracks, it is important to place constraints on
the population under consideration.  We cannot have an arbitrary
number of stars leaving the MSTO and eventually passing through the HB
phase; the bolometric luminosity of the population and the stellar
lifetimes constrain the number of stars in a given evolutionary phase
(Greggio \& Renzini 1990\markcite{GR90}; Renzini 1998\markcite{R98}).
The stellar death rate, or stellar evolutionary flux (SEF), is given
by the relation SEF $=B(t) L_T$, where $L_T$ is the total bolometric
luminosity of the population, and $B(t)$ is the specific evolutionary
flux.  The specific evolutionary flux is a weak function of age, and
for a population of age $\sim 10$ Gyr, $B(t) \cong 2.2\times 10^{-11}$
stars yr$^{-1}$~$L_{\odot}^{-1}$ (Renzini 1998\markcite{R98}).

The section of the STIS image most appropriate for an analysis of the
HB is the sum of regions 1, 2, and 3 (see Table \tabphot).  Restricting our
comparison to this deep portion of the STIS image avoids the
complications of crowding, systematic errors due to a steeply varying
background, and seriously incomplete photometry at the magnitudes of
interest ($\sim 27$~mag).  We will refer to this section as $R_{123}$;
the left and right edges of this section are 20$\arcsec$ and
8$\arcsec$ from the center of M32, respectively.  Later analyses,
especially if color information becomes available on these stars, may
include the full catalog, with some appropriately bright magnitude
cutoff in the regions that are very crowded.  The luminosity function
(LF) in $R_{123}$ is shown in Figure~\figehb.  Two key features of
this LF, which will be addressed later, are the lack of stars brighter
than 22 mag and the large number of stars near 26 mag.

We determined the bolometric luminosity in $R_{123}$ in the following
manner.  First, we registered the archival WFPC2 F555W image (HST
Guest Observer ID No. 5236) to our STIS field and summed the count
rate in $R_{123}$ to obtain 2.62$\times 10^4$ cts sec$^{-1}$.  The
corresponding $M_{F555W}$ is 11.44 mag on the STMAG system.
Conversion to Johnson V involved a small, color-dependent correction;
we used the M32 nuclear spectrum to derive an offset of $-0.03$ mag,
giving $m_V = 11.41$~mag.  Assuming a foreground extinction of
$E(B-V)=0.11$~mag with $R_V=3.1$ (see \S\ref{secobs}) and a distance
modulus of 24.43 mag gives $M_V=-13.36$~mag.  Worthey
(1994\markcite{W94}) gives bolometric corrections for passively
evolving populations as a function of age and metallicity; for these
purposes, we assume an age of 8~Gyr and solar metallicity (see
Grillmair et al.\ 1996\markcite{G96}; Da Costa 1997\markcite{D97} and
references therein), giving $BC_V=-0.875$~mag.  Note that this
parameter is sensitive to the assumed age and metallicity.  At
[Fe/H]~=~0.0 and an age of 5 Gyr or 12 Gyr, $BC_V$ is $-0.768$ or
$-0.963$~mag, respectively; at 8 Gyr and [Fe/H]~=~$-0.25$ or +0.25,
$BC_V$ is $-0.679$ or $-1.130$, respectively.  Using our assumed
metallicity and age, the bolometric luminosity is 3.92$\times 
10^7$~$L_{\odot}$ in $R_{123}$, and thus the associated SEF is $8.62\times
10^{-4}$ star yr$^{-1}$.  This is the upper limit we will place upon
the number of stars entering the evolutionary tracks in the following
discussion.  Note that the SEF in the entire STIS image, which
includes significantly more luminosity than that in $R_{123}$, is
$4.18\times 10^{-3}$ star yr$^{-1}$, using the same calculations and
assumptions.

In the subsequent discussion, we will translate tracks from the literature
and from our own calculations into magnitudes in the STIS bandpass.
These translations will be obtained from the solar-metallicity
Kurucz (1993\markcite{K93}) synthetic spectra, interpolating
in effective temperature from the grid points with the 
closest match in surface gravity and metallicity.  We then assume
a foreground reddening of $E(B-V) =0.11$~mag, following the
Cardelli, Clayton, \& Mathis (1989\markcite{CCM89}) parameterization,
and a distance of 770 kpc.  Because we are comparing to the STIS
data in 0.5~mag bins, our analysis is not very sensitive to these
assumptions.

\smallskip \hskip -0.1in
\parbox{3.5in}{\epsfxsize=3.5in \epsfbox{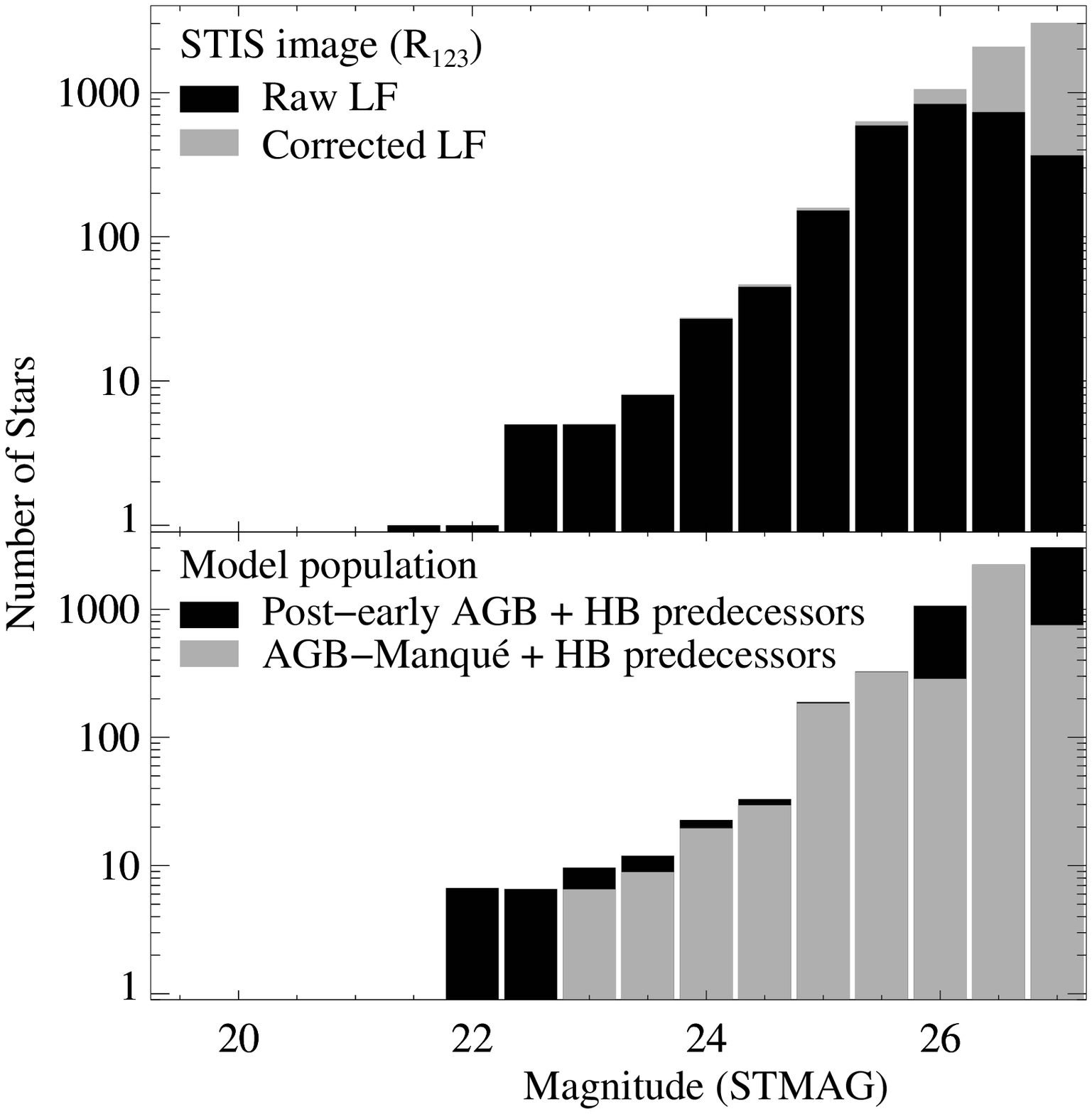}}\vskip 0.1in
\centerline{\parbox{3.0in}{\small {\sc Fig.~\figehb--} The STIS
luminosity function in the deepest half of the image ($R_{123}$) is
shown with and without corrections for incompleteness in the number
counts (top panel).  A minority population of hot HB stars and their
progeny (bottom panel; see \S\ref{secehb}) can easily explain the STIS
luminosity function.  This model population is comprised of the tracks
listed in Table \tabmod.  Note the logarithmic scaling.  }}
\addtocounter{figure}{1}

\medskip

\subsection{Post-AGB Evolution} \label{secpagb}

At an age of 8 Gyr, the main sequence turnoff mass is $\sim 1.1$~$M_{\odot}$ 
at solar metallicity (Bertelli et al.\ 1994\markcite{B94}),
and the core mass of post-AGB stars is expected to be low ($\lesssim
0.6$~$M_{\odot}$; Vassiliadis \& Wood 1993\markcite{VW93}).  Assuming
that the dominant population in M32 is at an age of at least 8~Gyr,
most of the stars that leave the red end of the HB should pass through
the post-AGB tracks of lowest mass in the compilations of Sch$\ddot{\rm
o}$nberner (1987\markcite{S87}) and Vassiliadis \& Wood
(1994\markcite{VW94}).  The luminosities of these stars as they cross
the HR diagram from the AGB to the white dwarf cooling curve are
listed in Table \tabplum.  We are comparing the STIS data to these
tracks from the literature because they have been widely used; our own set
of evolutionary models includes post-AGB behavior that produces very
similar luminosity functions.  Note that our post-AGB track shown in
Figure 1 crosses the HR diagram at a very low-mass (0.538 $M_\odot$),
and is thus somewhat dimmer in luminosity than the tracks taken from
the literature.

A post-AGB star should cross the HR diagram at a luminosity near that
of the AGB tip luminosity.  Several recent investigations of the AGB
tip luminosity in M32 show there is a significant population of stars
on the AGB at luminosities brighter than the tip of the red giant
branch (RGB) (see Grillmair et al.\ 1996 and references therein).
However, earlier studies that found very bright AGB stars, like those
associated with young populations, were apparently affected by
crowding; instead, it appears that the bright AGB stars in M32 belong
to a population of long period variables (LPVs) and blue-straggler
progeny, like those found in the old metal-rich globular clusters 47
Tuc and NGC6553 (Guarnieri, Renzini, \& Ortolani 1997\markcite{GRO97};
Renzini 1998\markcite{R98}; Grillmair et al.\ 1996\markcite{G96}).
Note that in a solar-metallicity population of age 8 Gyr, the RGB tip
lies at $log(L/L_{\odot}) = 3.2$ (Bertelli et al.\
1994\markcite{B94}).  The lack of bright AGB stars associated with
young populations supports the premise that the majority of the
post-AGB stars in M32 should be evolving along low-mass tracks (near
this luminosity).

\medskip

\parbox{3.5in}{
{\sc Table \tabplum:} Post-AGB Crossing Luminosities

\begin{tabular}{|l|r|}
\tableline
Mass & $log(L/L_{\odot})$ \\
\tableline
H-burning 0.546 $M_{\odot}^a$ & 3.2 \\
H-burning 0.569 $M_{\odot}^b$ & 3.5 \\
H-burning 0.597 $M_{\odot}^b$ & 3.7 \\
H-burning 0.633 $M_{\odot}^b$ & 3.9 \\
He-burning 0.567 $M_{\odot}^b$ & 3.5 \\
He-burning 0.600 $M_{\odot}^b$ & 3.6\\
\tableline
\end{tabular}

\noindent
$^a$Sch$\ddot{\rm o}$nberner (1987\markcite{S87}). A post-early AGB track 
just below\\ the mass required for post-AGB behavior.\\
$^b$Vassiliadis \& Wood (1994\markcite{VW94}).\\
}

\medskip

Until the middle of this decade, post-AGB stars were thought to be
predominantly H-burning (see Vassiliadis \& Wood 1994\markcite{VW94}
and references therein).  H-burning post-AGB stars leave the AGB
between He-shell flashes, while He-burning post-AGB stars leave the
AGB near or during a flash, and the ratio of He-burning to H-burning
lifetimes over the He-shell flash cycle was thought to be on the order
of 20\%.  However, Vassiliadis \& Wood (1994\markcite{VW94}) suggested
that at lower masses, the chance of producing a He-burning post-AGB
star increased (see also Renzini \& Fusi Pecci 1988\markcite{RF88};
Renzini 1989\markcite{R89}).  Post-AGB evolution remains as one of the
least-understood phases of normal stellar evolution; more 
observational evidence is needed to determine the relative frequency
of H-burning and He-burning tracks (see Dopita, Jacoby, \& Vassiliadis 
1992\markcite{DJV92}).\\

\hspace{0.5in}
\parbox{5.5in}{
{\sc Table \tabpagb:} Comparison with Post-AGB Tracks

\begin{tabular}{|c|r|r|r|r|r|r|r|r|}
\tableline
          & 
\multicolumn{2}{|c|}{Data} & 
\multicolumn{5}{|c|}{Theoretical Maximum Number of Post-AGB Stars} \\
\tableline
          & 
Raw LF & 
Corrected LF & 
0.546 $M_{\odot}^a$ & 
0.569 $M_{\odot}^b$ & 
0.597 $M_{\odot}^b$ & 
0.567 $M_{\odot}^b$ & 
0.600 $M_{\odot}^b$ \\
mag & 
regions 1+2+3 & 
regions 1+2+3 & 
H burning & 
H burning & 
H burning & 
He burning &
He burning \\
\tableline
20.0      & 0    & 0    &  0	&  0    &  1    &  0 &  0\\
20.5      & 0    & 0    &  0	&  6    &  2    &  1 &  0\\
21.0      & 0    & 0    &  0	&  4    &  0    &  1 &  2\\
21.5      & 1    & 1    &  54	&  3    &  1    &  1 &  1\\
22.0      & 1    & 1    &  40	&  2    &  1    &  1 &  4\\
22.5      & 5    & 5    &  30	&  2    &  1    &  2 & 14\\
23.0      & 5    & 5    &  23	&  2    &  0    &  2 &  2\\
23.5      & 8    & 8    &  24	&  2    &  1    & 29 &  2\\
24.0      & 27   & 27   &  29	&  2    &  1    &  4 &  1\\
24.5      & 45   & 47   &  26	&  2    &  1    &  4 &  1\\
25.0      & 152  & 159  &  26	&  2    &  0    &  3 &  1\\
25.5      & 590  & 631  &  16	&  1    &  1    &  3 &  1\\
26.0      & 832  & 1055 &  15	&  2    &  0    &  3 &  1\\
26.5      & 732  & 2079 &  22	&  7    &  1    & 11 &  4\\
27.0      & 366  & 3045 &  32	& 33    &  3    & 22 & 10\\
\tableline
\end{tabular}
$^a$Sch$\ddot{\rm o}$nberner (1987). A post-early AGB track 
just below the mass required for post-AGB behavior.\\
$^b$Vassiliadis \& Wood (1994).\\
}

M32 has a very weak UV upturn, and thus, prior to our observations,
the UV flux could theoretically be explained by low-mass post-AGB
stars alone; no hot HB stars were required.  We know that hot HB
stars, if present, can only comprise a minority of the population,
else the UV upturn in M32 would be considerably stronger; the
elliptical galaxies with the strongest UV upturns only require $\sim
10$\% of the SEF to pass through the hot HB and its descendants (Brown
et al.\ 1997\markcite{BFDD97}).  We expect most of the stars in M32 to
pass through the red end of the HB and the subsequent low-mass
post-AGB tracks.  While post-AGB stars evolve very rapidly through
their UV-bright phases compared to the hot HB stars and their progeny
(see \S\ref{secint}), the low-mass post-AGB tracks nevertheless evolve
slowly enough to produce dozens of stars at very bright magnitudes in
the STIS bandpass.  This is shown in Table \tabpagb, which shows the
luminosity functions obtained by placing the entire SEF (determined
for $R_{123}$ in \S\ref{secsef}) into selected low-mass tracks from
the literature.  For comparison, we show the raw and corrected
luminosity functions obtained in the deepest half of the STIS image,
$R_{123}$ (see \S\ref{secsef}).  It is obvious from Table \tabpagb\
that low-mass H-burning tracks suffer from two problems when trying to
explain the STIS luminosity function: the models predict far too many
stars at bright magnitudes ($\lesssim 23$ mag) and far too few stars
at faint magnitudes ($\gtrsim 24$ mag), even when the faint magnitudes
are not corrected for completeness.  Low-mass He-burning post-AGB
tracks fair somewhat better with the bright end of the STIS luminosity
function, as they do not predict nearly as many bright stars, but
these tracks still produce far too few stars at faint magnitudes.
These He-burning tracks also produce a fairly bright local maximum in
their luminosity functions (see Table~\tabpagb) that is not seen in
the STIS data, but the luminosity of this phase depends upon the mass
and the details of the evolution.  Thus this spike in the LF could be
easily hidden in the STIS data if there was a small dispersion in
post-AGB mass, centered at a low mass ($\lesssim 0.567$~$M_\odot$).
 
The lack of UV-bright stars in the STIS data requires that the
transition from the AGB to $T_{\rm eff} > 60000$~K occurs on a much
more rapid timescale than predicted by the low-mass H-burning tracks,
or, alternatively, that this transition be hidden by circumstellar
dust produced during the preceding AGB mass-loss (see

\newpage

\noindent
Figure~\figtran).  If the transition is only accelerated up to a
somewhat cooler temperature (e.g., 30000~K or 50000~K), there will
still be too many stars in the LF bins brighter than 23.5 mag.  The pace
of this transition is naturally increased by assuming post-AGB tracks
of higher mass ($\gtrsim 0.6$ $M_{\odot}$), but, as we stated above,
the observational and theoretical evidence on the maximum AGB stellar
luminosity seems to preclude this option.  Furthermore, the AGB
precursors of a population of stars this massive would produce an
enormous amount of energy, leading to optical-infrared colors in
disagreement with observations (Greggio \& Renzini
1999\markcite{GR99}).  Circumstellar extinction, while having a strong
effect in the UV, would also appear to be an unlikely candidate,
considering the rapid thinning times for material surrounding post-AGB
stars (K$\ddot{\rm a}$ufl, Renzini, \& Stanghellini
1993\markcite{KRS93}; Brown et al.\ 1998\markcite{BFSD98}), although
we cannot rule out this alternative.

\parbox{3.0in}{\epsfxsize=3.5in \epsfbox{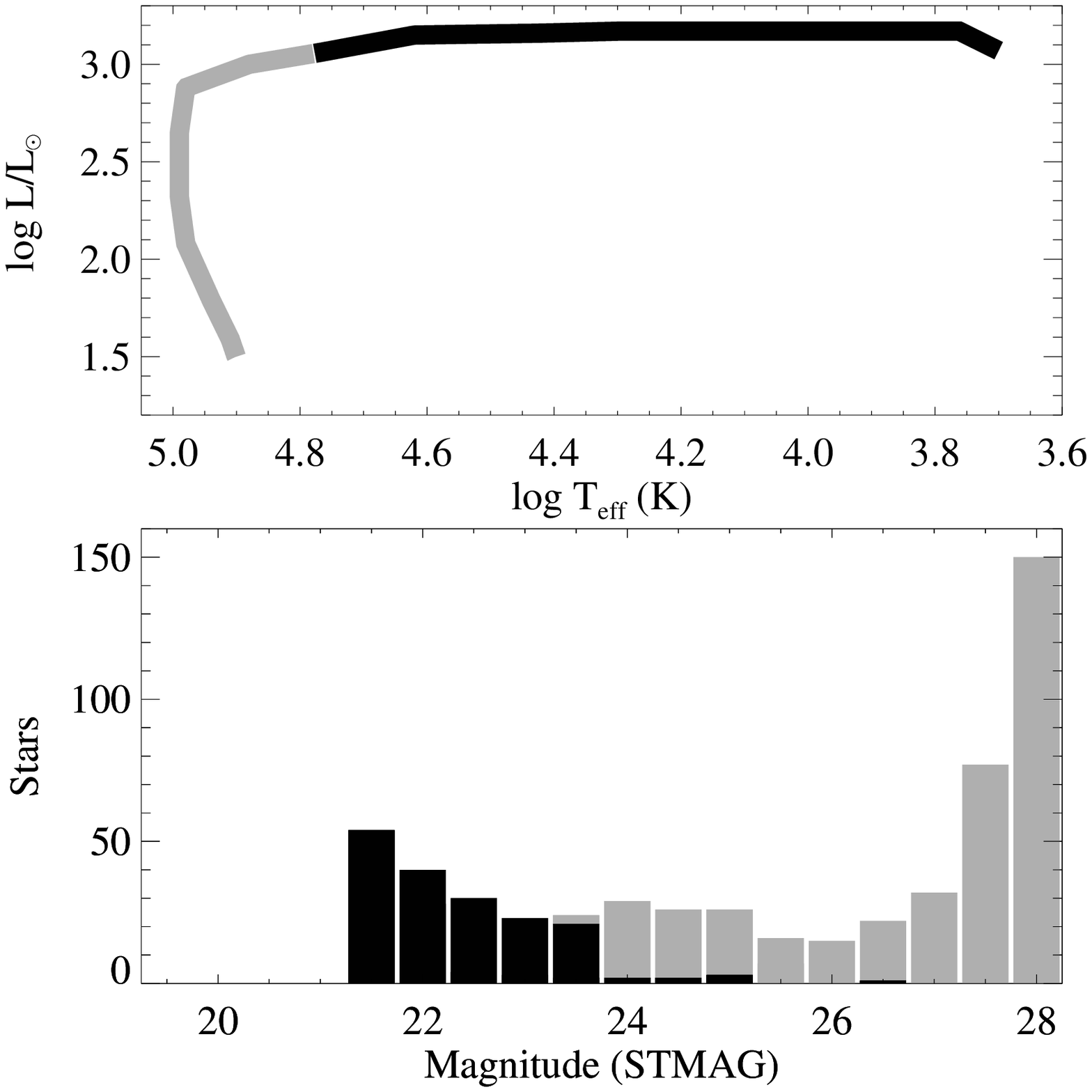}}
\centerline{\parbox{3.0in}{\small {\sc Fig.~\figtran--} The 0.546
$M_\odot$ H-burning track of Sch$\ddot{\rm o}$nberner (1987), and the
corresponding luminosity function from Table~\tabpagb.  The track and
model luminosity function have been color-coded to demonstrate that
the evolution up to $\rm T_{eff}$ = 60000~K causes the discrepancy
with the STIS luminosity function (i.e., the dark segment of the track
produces the dark portion of the luminosity function bins).  If this
evolution were more rapid or hidden, the lack of UV-bright stars in
the STIS image would be explained.  }} \addtocounter{figure}{1}
\smallskip

The most likely explanation for the lack of bright stars is that most
of the population evolves more rapidly from the AGB to high effective
temperature than predicted by the tracks in Tables \tabplum\ and
\tabpagb.  This might occur if the final mass ejection on the AGB was
triggered by a helium-shell flash which left the star out of thermal
equilibrium.  The post-AGB evolution would then take place on the more
rapid thermal timescale, as argued by K$\ddot{\rm a}$ufl et al.\
(1993\markcite{KRS93}) and Greggio \& Renzini (1999\markcite{GR99}).
Indeed, Table \tabpagb\ already demonstrates that the discrepancy with
the STIS data is greatly reduced if the stars evolve as He-burners.
Note that the number of known planetary nebulae in the STIS field
(four), combined with the stellar evolutionary flux for the entire
STIS field ($4.18\times 10^{-3}$ star yr$^{-1}$), implies that the
planetary nebula phase only lasts for $\sim 1000$~yr.  Although we
might have an explanation for the lack of bright stars, there is no
post-AGB track that can produce the thousands of stars on the faint
end of the STIS luminosity function, and so we next explore the
expectations from hot horizontal branch stars and their progeny.

\subsection{Hot HB stars} \label{secehb}

As explained in \S\ref{secint}, the HB phase (shown as a light grey
region in Figures \figtrka\ and \figtrkb) lasts for $\sim10^8$ yr, and
the UV-bright post-HB lifetime for stars leaving the hot end of the HB
is on the order of $10^7$ yr.  These long lifetimes in the UV suggest
that hot HB stars and their progeny are much more likely to explain
the STIS luminosity function, compared to the brighter short-lived post-AGB
stars.  We explore this possibility here, using our own calculations
of HB and post-HB evolution.

\subsubsection{Models} \label{secmod}

We have calculated a detailed grid of HB and post-HB sequences with
solar metallicity and helium abundance, over a wide range in the HB
mass, and for various rates of mass loss along the AGB.  A fine
spacing in the HB mass was used, in order to clearly define the
changes in the HB morphology with mass, as well as the transition between the
AGB-Manqu$\acute{\rm e}$, post-early-AGB, and post-AGB evolution.
Evolution was followed from the ZAHB until the luminosity fell below
0.1 $L_{\odot}$ along the white dwarf cooling curve.  In a few cases,
the models underwent a final helium-shell flash while descending the
WD cooling curve, and such sequences were stopped if the flash
convection reached into the hydrogen envelope.

In order to obtain a ZAHB model at the red end of the HB, we first evolved
a 1 $M_{\odot}$ stellar model from the zero-age main sequence up the
RGB, and then through the helium-core flash.  
The parameters of the initial
main-sequence model were determined by calibrating the model on the sun;
i.e., the main sequence helium abundance $Y_{MS}$, heavy-element
abundance $Z$,
and mixing-length ratio $\alpha$ were adjusted until the
model matched the observed solar luminosity, radius, and $Z/X$ ratio at
a solar age of 4.6 Gyr.  The parameters derived in this fashion are:
$Y_{MS} =0.2798$, $Z=0.01716$, and $\alpha = 1.8452$.  The helium abundance
increased to 0.3003 during the first dredge-up along the lower RGB.  
Consequently, all of the HB and AGB sequences we have computed have
the following envelope abundances:
$Y=0.3003$ and $Z=0.01716$.

Mass loss was included during the RGB phase using the Reimers 
(1975\markcite{R75}) mass-loss formulation, with a mass-loss
parameter $\eta_R$ of 0.4.  As a result, the mass decreased 
to 0.8269 $M_{\odot}$ by the time the model reached the
ZAHB.  The age at that time was 12.2 Gyr.  The use of a somewhat different
initial mass (corresponding to a somewhat different age) would have only
a negligible effect on the core mass and envelope helium abundance of this
ZAHB model.  Thus our results do not depend significantly on the
choice for the zero-age main sequence model.

Lower mass ZAHB models were then computed by removing mass from the
envelope of the red ZAHB model described above.  The
lowest ZAHB mass considered here was 0.473 $M_{\odot}$, corresponding
to an envelope mass of only 0.00162 $M_{\odot}$.

The ZAHB models computed in this manner were then evolved through the
HB phase using standard algorithms for convective overshooting and
semiconvection.  The last model from each HB sequence was then used as
the starting model for the subsequent AGB evolution.  Mass loss was
included in the AGB sequences, using the Reimers (1975\markcite{R75})
formulation for three values of $\eta_R$: 0.0 (no mass loss), 0.4, and
1.0.  Note that these assumptions for AGB mass loss have no
consequence for the AGB-Manqu$\acute{\rm e}$ evolution, because such
stars do not ascend the AGB (see \S\ref{secint}).  For this reason,
and because we lack color information for these stars (given the one
bandpass of our observations), we will only explore the models that
assume $\eta_R = 0.4$ in our discussion below.  The variation in mass
loss may be explored in future work if color information becomes
available for these stars.

The set of tracks with $\eta_R = 0.4$ consisted of 35 ZAHB masses
(M$_{ZAHB}$), ranging from 0.473 to 0.700 $M_{\odot}$ and covering a
temperature range from $log \rm T_{eff} =$ 4.42 -- 3.67.  Due to mass
loss on the AGB, the final masses on the white dwarf cooling curve
($M_{WD}$) for these tracks is somewhat less.  AGB-Manqu$\acute{\rm e}$
behavior was found for stars with $M_{ZAHB} \le 0.505$~$M_{\odot}$;
post-early AGB behavior for stars with $0.505 < M_{ZAHB} <
0.610$~$M_{\odot}$ ($0.505 < M_{WD} < 0.542$~$M_{\odot}$), and
post-AGB behavior for stars with $M_{ZAHB} \ge 0.610$~$M_{\odot}$
($M_{WD} \ge 0.542$~$M_{\odot}$).

These tracks permit the
transition between the different classes of post-HB evolution
to be much better defined than previously possible.
Figure \figtrkb\ shows examples of these evolutionary tracks
as they appear in the STIS bandpass (see also Figure \figtrka).

\parbox{3.0in}{\epsfxsize=3.5in \epsfbox{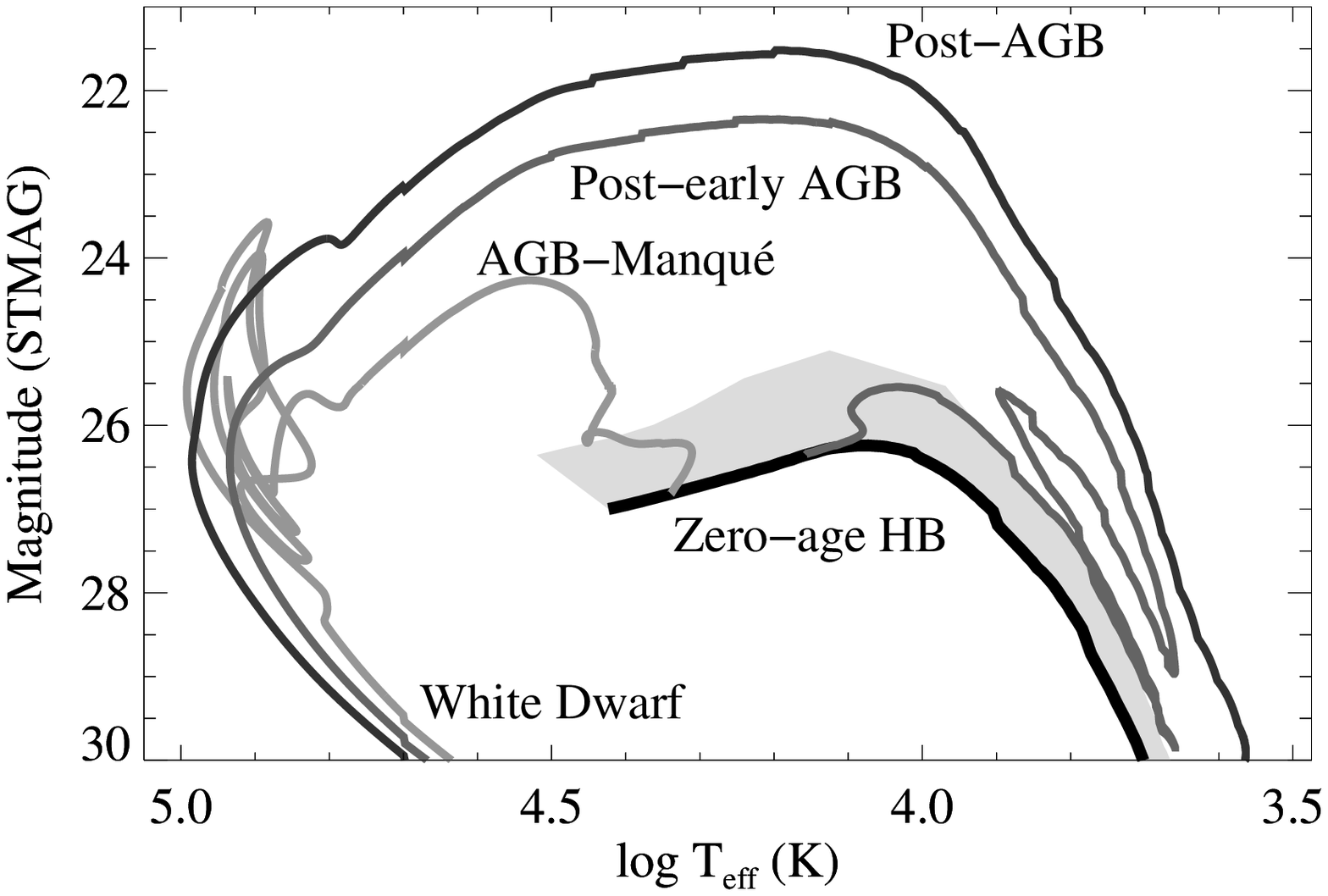}}
\centerline{\parbox{3.0in}{\small {\sc Fig.~\figtrkb--}
The same tracks shown in Figure 1, but with luminosity replaced by
magnitude in the STIS bandpass, under the STMAG system, assuming a distance
of 770 kpc and $E(B-V) = 0.11$ mag.  Note that assuming a 
lower reddening of $E(B-V) = 0.08$ mag would make the tracks
$\sim 0.25$ mag brighter.  The HB phase
(again shown by light grey shading, as determined from our entire
set of tracks), spans a range of 25--27 mag at $T_{\rm eff} > 8500$~K.
}}
\addtocounter{figure}{1}
\smallskip

\subsubsection{Hot HB Luminosity Functions}

We have constructed the luminosity functions for each of the above HB
and post-HB tracks in the STIS bandpass for comparison with the
corrected STIS luminosity function given in Table \tabpagb.  Because
we only have one bandpass and no color information, it is difficult to
constrain the mass distribution of stars on the hot HB.  This
difficulty is further compounded by the fact that post-AGB stars
certainly contribute some small but non-negligible component to the
STIS luminosity function, because we know the majority of the evolved
population passes through some sort of post-AGB phase, given the weak
UV upturn in M32.  Color information would also help to separate these
post-AGB stars from the rest of the STIS data, because they spend most
of their time at much higher effective temperature.

The luminosity functions for four of our tracks can be added to create
a composite luminosity function that agrees quite well with the STIS
data (Figure \figehb).  Two of these tracks follow
AGB-Manqu$\acute{\rm e}$ evolution and two follow post-early AGB
evolution upon leaving the HB.  These tracks have been normalized to
maximize agreement with the STIS data from 23--27 mag, and we show in
Table \tabmod\ the relative contributions of these tracks to this
composite LF.  We stress that this LF serves as an example, to show
that hot HB stars can reproduce the STIS luminosity function; color
information is required to properly constrain the true distribution of
mass on the hot HB.  The STIS data confirm that stars passing through
the hot HB at $T_{\rm eff} > 8500$~K comprise only a small fraction
(approximately 7\%) of the total HB population in M32.  This fraction
of hot HB stars would be somewhat lower if the contribution of the
post-AGB stars could be subtracted from the STIS luminosity function.

Given the small fraction of the population entering the hot HB, we
must conclude that the vast majority of the population passes through
the red HB phase and the subsequent post-AGB evolution.  Although a
small population of hot HB stars can explain the numerous faint stars
present in the STIS image, we still must explain the lack of bright
stars.  The likely explanation, as discussed in \S\ref{secpagb}, is
that the transition time from the AGB to the hotter post-AGB phases
must be more rapid than that expected from the canonical low-mass
H-burning tracks.  The contribution of post-AGB stars to the STIS
luminosity function would then be small, even though most of the population
channels through post-AGB tracks. 

\medskip

\hskip 0.25in
\parbox{3.5in}{
{\sc Table \tabmod:} Hot HB Components of Composite LF

\begin{tabular}{|r|r|r|r|}
\tableline
$M_{ZAHB}$     & ZAHB &SEF               & Fraction of \\
($M_{\odot}$)  & $T_{\rm eff}$ (K) & (stars yr$^{-1}$) & total SEF \\

\tableline
0.475 & 24578 & $1.95\times10^{-5}$ & 0.023 \\ 
0.500 & 16372 & $3.03\times10^{-6}$ & 0.004 \\ 
0.520 & 12034 & $6.20\times10^{-6}$ & 0.007 \\ 
0.545 & 6057  & $3.20\times10^{-5}$ & 0.037 \\ 

\tableline
\end{tabular}
}

\smallskip

\section{DISCUSSION}

\subsection{Dynamical Creation Mechanism for Hot HB Stars?}

In the Galactic field, hot HB and post-HB stars are often found in
binaries, and it has been suggested that a dynamical mechanism may
play a role in the production of hot HB stars in old, metal-rich
populations such as elliptical galaxies (Green et al.\
1997\markcite{G97}; Green \& Chaboyer 1998\markcite{GC98}).  In
globular clusters, the role of dynamics remains a matter of debate.
Horizontal branch morphology tends to become redder as metallicity
increases, but examples of metal-rich GCs with blue HB morphology
demonstrate that other parameters are at work (e.g., Rich et al.\
1997\markcite{R97}).  One of these parameters may be a dynamical
mechanism, through binary or tidal interaction while on the RGB.
There is some evidence that dynamics plays a role in HB morphology;
e.g., Fusi Pecci et al.\ (1993\markcite{F93}) and Buonanno et al.\
(1997\markcite{B97}) demonstrated that more concentrated clusters have
bluer HB morphologies with extended blue tails.  Rich et al.\
(1997\markcite{R97}) found two metal-rich GCs with extended HB
morphology (NGC6388 and NGC6441), although the evidence for a
dynamical origin in that work appears inconclusive.  These clusters
have some of the highest values of central surface brightness,
velocity dispersion, and stellar collision rates for Galactic GCs,
which may partially explain their blue HBs, but Rich et al.\
(1997\markcite{R97}) find no evidence for a difference in the radial
distribution of blue and red HB stars, as expected if tidal or binary
interaction plays a role.  Later work by Layden et al.\
(1999\markcite{L99}) did find that the blue HB stars in NGC6441 were
more centrally concentrated than the red HB stars, but the difference
in these gradients does not conform with theoretical expectations for
a dynamical mechanism, because it occurs too far from the cluster
center.  In NGC6752, Landsman et al.\ (1996\markcite{L96}) found that
the hot HB is a continuous extension of the intermediate-temperature
HB population, suggesting that a common single-star mechanism is
responsible for both.  In $\omega$ Cen, which has the largest known
fraction of hot HB stars, D'Cruz et al.\ (1999\markcite{D99}) find no
evidence for a radial gradient in the hot HB to red HB number ratio.

M32 is much smaller than other well-studied true ellipticals, and it
has the lowest Mg$_2$ optical metallicity index of all the quiescent
UV upturn galaxies in the Burstein et al.\ (1988\markcite{B88})
sample; in this sense, it is the elliptical galaxy that lies closest
to the realm of globular clusters.  However, elliptical galaxies
should not be considered as overgrown globular clusters -- they
clearly inhabit a different regime of parameter space, especially when
considering the evolved population of stars that produces the UV
upturn.  This is apparent in the modified version of the Burstein et
al.\ (1988\markcite{B88}) $1550-V$ vs. Mg$_2$ relation shown by Dorman
et al.\ (1995\markcite{DRO95}), which shows the GCs lying in a
completely distinct clump removed from the tight E galaxy sequence.
It is thus worth noting that the existence of hot HB stars in M32
cannot be easily explained by dynamical mechanisms, because the
stellar densities in elliptical galaxies are so much lower than those
in globular clusters (except for the very center of ellipticals, which
often harbor a black hole).  The central luminosity densities of
NGC6388 and NGC6441 are respectively $1.86 \times 10^5$~$L_{\odot}$
pc$^{-3}$ and $2.00 \times 10^5$~$L_{\odot}$ pc$^{-3}$ (Djorgovski
1993\markcite{D93}).  M32 has a luminosity density of $4.9 \times 
10^5$~$L_{\odot}$ pc$^{-3}$ at a radius of $0.1 \arcsec$, but this density
drops rapidly by more than three orders of magnitude at radii greater
than 8$\arcsec$ (Tonry 1988\markcite{To88};
Gebhardt et al.\ 1996\markcite{GRA96}; private communication Gebhardt
1999).  Furthermore, the far-UV to $B$-band flux ratio increases with
radius in M32 (Ohl et al.\ 1998\markcite{O98}).  Because our STIS data
firmly demonstrate that the weak far-UV flux is due to hot HB stars,
the increase in the far-UV to $B$-band flux ratio indicates that hot
HB stars comprise a larger fraction of the population at increasing
distance from the galactic center -- again, opposite to the behavior
expected from a dynamical creation mechanism such as tidal interaction
or binarism. 

\subsection{The Age of M32}

In recent years, much of the effort on dating M32 has shown a need for
an intermediate-age ($\sim 5$ Gyr) component to its population.  When
this fact is considered in tandem with its small size and proximity to
M31, its utility as a template for other ellipticals is somewhat
diminished.  However, even though it may be unusual, M32 represents
the only elliptical near enough for intense study by all of the
standard stellar population analysis methods (spectroscopy, line
indices, color-magnitude diagrams, etc.), at least until we have a
successor to HST with UV-optical capability.  With this in mind, are
we certain that M32 has a younger component to its older, dominant
population?  We feel that this remains an open question.  

Early color-magnitude diagrams reported evidence for a younger
population by finding very bright AGB stars, but later interpretations
of these data showed that both crowding (Renzini 1998\markcite{R98})
and long-period variables (Guarnieri et al.\ 1997\markcite{G97})
confused the earlier analysis; later observations (Grillmair et al.\
1996\markcite{G96}) also find no evidence for optically bright AGB
stars, but these stars were difficult to detect in the $V$ band.
Instead, it appears that the brightest AGB stars in M32 are consistent
with those found in other old globular clusters (Guarnieri et al.\
1997\markcite{G97}; Renzini 1998\markcite{R98}).  We note that our own
data show a lack of the very bright post-AGB stars that should be
present if they are leaving the tip of the AGB at luminosities much
brighter than the RGB tip.

Spectral synthesis analyses using isochrones like those of Worthey
(1994\markcite{W94}) assume a pure red clump HB morphology, and the
STIS data show that this assumption is not entirely accurate; note,
however, that the Worthey models were not intended to address UV flux.
Depending upon how much of the population extends beyond the red
clump, an analysis that assumes a pure red clump may be seriously
incorrect.  Our data demonstrate that the hot horizontal branch is
populated in M32, but the lack of color information prevents us from
constraining the HB morphology further.  The tracks that reproduce the
STIS luminosity function come from a wide range of T$_{\rm eff}$ on
the HB (see Table \tabmod), and so they are consistent with both a
bimodal HB distribution and with an extended, more uniform
distribution.  However, the integrated spectrum of M32 does not show
the strong 2500 \AA\ dip seen in UV-bright quiescent giant ellipticals
(compare Figure \figspec\ with the spectra of Burstein et al.\
1988\markcite{B88} and Brown et al.\ 1997\markcite{BFDD97}).  Given
the presence of hot HB stars and the missing 2500 \AA\ dip, the
distribution of effective temperature on the HB might be more uniform
in M32 than the strongly bimodal distributions assumed in giant
ellipticals.  The zero-age HB becomes more bimodal at increasing
metallicity and helium abundance (see Dorman et al.\
1993\markcite{DRO93}); because M32 has an abundance much closer to
solar than the giant ellipticals (see Burstein et al.\
1988\markcite{B88}), evolution theory favors a more uniform HB.  Our
hot HB model population (Table~\tabmod), if correct, would be
responsible for practically all of the flux at 2500~\AA, and
contribute to approximately 10\% of the flux from 3000--4000~\AA, even
though it only comprises a small fraction of the HB population
($\sim$5\%).  Thus, this hot HB component should not be ignored for
population fitting in the mid-UV.

The H$_\beta$ ($\lambda$ 4861~\AA) line index is often quoted as
another piece of evidence in favor of a young component in M32.  It is
significantly stronger in M32 than in the higher metallicity giant
ellipticals (Bressan, Chiosi, \& Tantalo 1996\markcite{BCT96};
Burstein et al.\ 1984\markcite{BFGK84}), with an equivalent width (EW)
of 2.2~\AA\ compared to an average value of 1.7~\AA.  Metal-poor
globular clusters have even stronger H$_\beta$ absorption than M32 (as
high as 3~\AA\ EW), but the index decreases in strength at increasing
metallicity, and high metallicity globular clusters show a weaker
index than in M32 (Burstein et al.\ 1984\markcite{BFGK84}).  Note,
however, that these earlier studies did not include metal-rich GCs
with blue HB morphology, such as those studied by Rich et al.\
(1997\markcite{R97}).  Burstein et al.\ (1984\markcite{BFGK84}) claim
that the addition of early A stars to the old stellar population in
M32 can increase the H$_\beta$ appropriately, but the resulting UV
continuum would be too bright, and instead favor later main-sequence F
stars.  Note that there is a radial gradient to the H$_\beta$
absorption in M32, in the sense that it is significantly weaker in the
deepest parts of the STIS image, compared to the galactic center
(Gonz$\acute{\rm a}$lez 1993\markcite{G93}; Hardy et al.\
1994\markcite{H94}); this suggests that the younger population, if
present, is most prevalant in the nucleus (see Grillmair et al.\
1996\markcite{G96}).

It is possible instead that intermediate-temperature HB stars could
account for {\it some} of the extra H$_\beta$ absorption (assuming
that M32 has a less bimodal HB mass distribution than seen in giant
ellipticals), but it is unlikely, given our data and models, that the
HB could account for {\it all} of this extra absorption.
Approximately 20\% of the light at H$_\beta$ comes from the horizontal
branch (see, e.g., Yi, Demarque, \& Oemler 1997\markcite{Y97}).  Thus,
if a large fraction of the ZAHB population was at F-star temperatures
(near 7000~K), the H$_\beta$ absorption would increase significantly.
Specifically, a small set of synthetic spectra calculated for this
purpose, using SYNSPEC (Hubeny, Lanz, \& Jeffery 1994\markcite{HLJ94})
and Kurucz (1993\markcite{K93}) model atmospheres, shows that the
H$_\beta$ EW increases from 1.2~\AA\ at T$_{\rm eff}$ = 5000~K to
8~\AA\ at 7500~K (measuring EW as in Faber et al.\
1985\markcite{F85}), eventually peaking to 9.5~\AA\ at 9000~K.  ZAHB
stars near 7000~K would lie at the very faint (and incomplete) end of
the STIS luminosity function (see Figure \figtrkb), and thus could be
present in large numbers.  However, these stars become post-early AGB
stars in their later and brighter phases, and if such tracks are
populated at significant levels ($\sim$ 50\%), the STIS luminosity
function would have many more UV-bright stars than we actually see.
Note that the hot-HB model population shown in Table~\tabmod\,
comprising $\sim$ 5\% of the HB population, would only contribute to
$\sim$ 5\% of the flux at H$_\beta$.  One could place a large fraction
of the ZAHB near 7000~K (thus accounting for all of the H$_\beta$
absorption) only if the later, UV-bright post-early-AGB phases were
more rapid than predicted from the models, as seen for the post-AGB
stars.  Thus, even with a hot HB component in M32, the H$_\beta$
absorption might be the one piece of evidence that is difficult to
explain without some trace of younger stars or blue stragglers.

\subsection{Summary}

The STIS data presented here are the first to directly image stars on
the hot horizontal branch in any elliptical galaxy.  In previous work,
the spectral energy distribution and the magnitude of the far-UV flux
in giant ellipticals required the presence of hot HB stars, unless
UV-bright post-AGB stars were much more efficient UV emitters than
thought previously.  Until these STIS observations, M32 was the one
example of an elliptical where the weak UV flux could have been
explained by canonical low-mass post-AGB tracks.  Our data show that the
hot HB is populated in M32, and that the UV-bright phases of post-AGB
evolution are less populated than expected from canonical tracks; thus
these data represent a direct confirmation that the UV upturn in
ellipticals originates in hot HB stars.  Our findings demonstrate that
M32 does not have a pure ``red clump'' HB morphology, as assumed by
many of the stellar population analyses of this galaxy.  If the HB
effective temperature distribution is not extremely bimodal, our
findings may weaken the evidence for an intermediate age population in
M32.  Color information would best constrain the HB morphology
further, and we will propose to carry out far-UV imaging of this same
field in the coming HST cycle; the far-UV data would provide an
excellent discriminator between stars hotter and cooler than 12000~K,
which would constrain the bimodality of the HB in M32.  Furthermore,
color information will allow us to discern how many of the faint stars
in our luminosity function are on the white dwarf cooling curve, and
this may shed light on the evolution of post-AGB stars, which appear
to be evolving rapidly in the STIS field.

\acknowledgments
Support for this work was provided by NASA through the STIS GTO team funding.
TMB acknowledges support at Goddard Space Flight Center by NAS~5-6499D.  
We wish to thank K. Gebhardt for kindly providing 
the luminosity density profile of M32.  We also wish to thank
P. Stetson, who provided the DAOPHOT-II package and gave us assistance
with its use.  This research has made use of the SIMBAD database, 
operated at CDS, Strasbourg, France.

\clearpage


\end{document}